\RequirePackage{lineno} 
\documentclass[preprint]{aastex}

\usepackage{natbib}
\usepackage{times}
\usepackage{hyperref}
\usepackage{longtable}

\slugcomment{to appear in the Astronomical Journal 
(submitted 25 May 2011)}

\newcommand{\mpg}{{g}}
\newcommand{\mpr}{{r}}
\newcommand{\mpi}{{i}}
\newcommand{\NCKB}{121 (64\%)}
\newcommand{\NRes}{58 (31\%)}
\newcommand{\NPlutino}{25 (13\%)}
\newcommand{\NScat}{9 (5\%)}
\newcommand{\Ninn}{6 (3\%)}
\newcommand{\Nmain}{101 (54\%)}
\newcommand{\Nout}{3 (2\%)}
\newcommand{\Ndetc}{11 (6\%)}
\newcommand{\NKBO}{196\ }
\newcommand{\lKBO}{231\ }
\newcommand{\TKBO}{169\ }

\newcommand{\AU}{{\rm AU}}

\shorttitle{CFEPS Full Data Release}
\shortauthors{Petit {\em et al.}}

\setlongtables

\begin{document}
\linenumbers

\title{The Canada-France Ecliptic Plane Survey - Full Data Release: 
\\ The orbital structure of the Kuiper belt\altaffilmark{1}}

\author{J-M. Petit\altaffilmark{2,4},
J.J. Kavelaars\altaffilmark{3},
B.J. Gladman\altaffilmark{4},
R.L. Jones\altaffilmark{3,4}, 
J.Wm. Parker\altaffilmark{5}, 
C. Van Laerhoven\altaffilmark{4,6},
P. Nicholson\altaffilmark{7},
G. Mars\altaffilmark{8},
P. Rousselot\altaffilmark{2},
O. Mousis\altaffilmark{2},
B. Marsden\altaffilmark{9,12}
A. Bieryla\altaffilmark{5},
M. Taylor\altaffilmark{10},
%I. Murray\altaffilmark{3},
M.L.N. Ashby\altaffilmark{9},
P. Benavidez\altaffilmark{11},
A. Campo Bagatin\altaffilmark{11},
G. Bernabeu\altaffilmark{11}
}

\altaffiltext{1}{Based on observations obtained with MegaPrime/MegaCam, a joint
  project of CFHT and CEA/DAPNIA, at the Canada-France-Hawaii Telescope (CFHT)
  which is operated by the National Research Council (NRC) of Canada, the
  Institute National des Sciences de l'Universe of the Centre National de la
  Recherche Scientifique (CNRS) of France, and the University of Hawaii. This
  work is based in part on data products produced at the Canadian Astronomy
  Data Centre as part of the Canada-France-Hawaii Telescope Legacy Survey, a
  collaborative project of NRC and CNRS.}
\altaffiltext{2}{Institut UTINAM, CNRS-UMR 6213, Observatoire de Besan\c{c}on,
  BP 1615, 25010 Besan\c{c}on Cedex, France}
\altaffiltext{3}{Herzberg Institute of Astrophysics, National Research
  Council of Canada, Victoria, BC V9E 2E7, Canada}
\altaffiltext{4}{Department of Physics and Astronomy, 6224
  Agricultural Road, University of British Columbia, Vancouver, BC, Canada}
\altaffiltext{5}{Planetary Science Directorate, Southwest Research
  Institute, 1050 Walnut Street, Suite 300, Boulder, CO 80302, USA}
\altaffiltext{6}{Department of Planetary Sciences, University of Arizona, 1629
  E. University Blvd, Tucson, AZ, 85721-0092, USA}
\altaffiltext{7}{Cornell University, Space Sciences Building, Ithaca, New York
  14853, USA}
\altaffiltext{8}{Observatoire de la Cote d'Azur, BP 4229, Boulevard de
  l'Observatoire, F-06304 Nice Cedex 4, France}
\altaffiltext{9}{Harvard-Smithsonian Center for Astrophysics, 60 Garden
  Street, Cambridge, MA 02138}
\altaffiltext{10}{Department of Physics and Astronomy, University of Victoria,
  Victoria, BC V8W 2Y2, Canada}
\altaffiltext{11}{Departamento de Fisica, Ingenieria de Sistemas y Teoria de la
  Se\~{n}al, E.P.S.A., Universidad de Alicante, Apartado de Correos 99,
  Alicante 03080, Spain}
\altaffiltext{12}{Deceased}

\begin{abstract}
We report the orbital distribution of the trans-neptunian objects (TNOs)
discovered during the Canada-France Ecliptic Plane Survey (CFEPS),
whose discovery phase ran from early 2003 until early 2007. The follow-up
observations started just after the first discoveries and extended until late
2009.
We obtained characterized observations of 321~sq.deg.~of sky to depths in 
the range $\mpg \sim$23.5~-~24.4~AB mag.
We provide a database of \TKBO\ TNOs with high-precision dynamical
classification and known discovery efficiency.
Using this database, we find that the classical belt is a complex region with
sub-structures that go beyond the usual splitting of inner (interior to 3:2
mean-motion resonance [MMR]), main (between 3:2 and 2:1 MMR), and outer
(exterior to 2:1 MMR). The main classical belt ($a$=40--47~AU) needs to be
modeled with at least three components: the `hot' component with a wide
inclination distribution and two `cold' components (stirred and kernel) with much narrower
inclination distributions.
The hot component must have a significantly shallower absolute magnitude ($H_g$)
distribution than the other two  components.
With 95\% confidence, there are $8000^{+1800}_{-1600}$ objects in the main belt
with H$_g \le 8.0$, of which 50\% are from the hot component, 
40\% from the stirred component and 10\% from the kernel; 
the hot component's fraction drops rapidly with increasing $H_g$.
Because of this, the apparent population fractions depend on the 
depth and ecliptic latitude of a trans-neptunian survey.
The stirred and kernel components are limited to only a portion of the 
main belt, while we find that the hot component 
is consistent with a smooth extension
throughout the inner, main and outer regions of the classical belt;  
in fact, the inner and outer belts are consistent with containing only 
hot-component objects. 
The $H_g \le 8.0$ TNO population estimates are $400$ for the inner belt and
10,000 for the outer belt  to within a factor of two (95\% confidence).
We show how the CFEPS Survey Simulator can be used to compare a cosmogonic model
for the the orbital element distribution to the real Kuiper belt.

\end{abstract}

\keywords{Kuiper Belt, surveys; PACS  96.30.Xa}

\hfill
\pagebreak

\section{Introduction}
The minor body populations of the solar system provide, via their orbital and
physical properties, windows into the dynamical and chemical history of the
Solar System.  Recognition of the structural complexity in the trans-neptunian
region has lead to models that describe possible dynamical evolutionary paths,
such as a smooth migration phase for Neptune \citep{1993Natur.365..819M}, the
large scale re-ordering of the outer solar system
\citep{2005Natur.435..459T,1999Natur.402..635T}, 
the scattering of now-gone rogue planets \citep{2006ApJ...643L.135G}, or the
close passage of a star \citep{2000ApJ...528..351I}.
Evaluating these models is fraught with
dangers due to observational biases affecting our knowledge of the intrinsic
populations of the trans-neptunian region \citep[see][for discussion of these
  issues]{2008ssbn.book...59K,2010AJ....139.2249J}.
Over the past twenty years, many different Kuiper Belt surveys 
(those with more
than 10 detections include
\citet{1996AJ....112.1225J,2001AJ....121..562L,2001AJ....122..457T,2001AJ....122.1051G,2002AJ....124.2949A,2002AJ....123.2083M,2005AJ....129.1117E,2006MNRAS.365..429P,2006Icar..185..508J,2010ApJ...720.1691S})
have been slowly building up a sample, albeit with differing
flux and pointing biases.
\citet{2006Icar..185..508J} 
enumerates the aspects of surveys that must be carefully recorded and
made public if quantitative comparisons with models are to be made.

The primary goal of the Canada-France Ecliptic Plane Survey (CFEPS) is the
production of a catalogue of trans-neptunian objects (TNOs) combined with a
precise account of the observational biases inherent to that catalog.  The
description of the biases, combined with provisioning of a `survey simulator',
enables researchers to quantitatively compare the outcome of their model
simulations to the observed TNO populations.
In \cite{2006Icar..185..508J}
we described our initial `pre-survey' and general motivation for this project,
and \cite{2009AJ....137.4917K} (P1 hereafter) describes the first year of
operation of this survey (the L3 data release).  
This manuscript describes the observations that make up the integrated
seven years of the project and provide our
complete catalog (the L7 release) of near-ecliptic detections and 
characterizations along with fully-linked high-quality orbits.
In summary, the `products' of the CFEPS survey consists of four items:
\begin{enumerate}
\item A list of detected CFEPS TNOs, associated with the block of discovery,
\item a characterization of each survey block, 
\item a Survey Simulator that takes the a proposed Kuiper Belt model, exposes
  it to the known detection biases of the CFEPS blocks and produces simulated
  detections to be compared with the real detections, and
\item the CFEPS-L7 model population.
\end{enumerate}

In Sections~2 and 3, we describe the observation and characterization of the
CFEPS TNO sample. The dynamical classification of all tracked TNOs in our
sample is given in Section~4. In Section~5, we update our parametrized model of
the main and inner classical Kuiper Belt (P1) and give an improved estimate of
the total number of objects in each of these dynamical subpopulations.
We also extend our model to the non-resonant, non-scattering part of the belt
beyond the 2:1 MMR with Neptune. Section~6 gives an order of magntitude
estimate of the scattering disk's population. Section~7 demonstrates
the use of our Survey Simulator to compare the results of a cosmogonic model to
the CFEPS detections. 
Finally in Section~8, we present our conclusions and put our
findings in perspective.

\section{Observations and Initial reductions}
\label{sec:obs}

The discovery component of the CFEPS project imaged $\sim$320 square degrees of
sky, almost all of which was within a few degrees of the ecliptic plane. 
Discovery observations occurred in blocks of $\approx16$ fields acquired using 
the Canada-France-Hawaii Telescope (CFHT) MegaPrime camera which delivered
discovery image quality (FWHM) of 0.7 - 0.9 arc-seconds in queue-mode
operations.
The $0.96^\circ \times 0.94^\circ$ MegaPrime FOV is paved by 36 individual 4600x2048 CCDs, each pixel having a scale of $0.187''$.  

\begin{figure*}[h]
\begin{center}
\includegraphics[width=16cm]{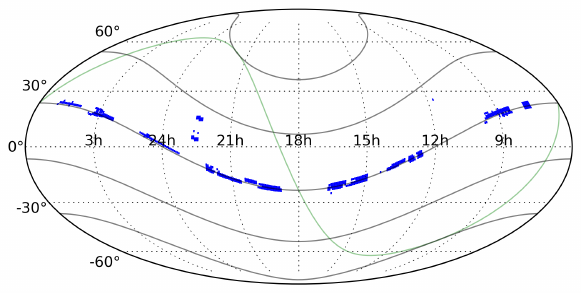}
\end{center}
\caption{Geometry of the CFEPS discovery-blocks. The RA and DEC
  grid is indicated with dotted lines. The black solid curves show
  constant ecliptic latitudes of -60$^\circ$, -30$^\circ$, 0$^\circ$,
  30$^\circ$, 60$^\circ$, from bottom to top. The remaining curve
  indicates the plane of the Milky Way.
}
\label{fig:cfeps_geom}
\end{figure*}

The CFEPS designation of a `block' of discovery fields was: 
a leading `L' followed by the year of observations (3,4,5 and 7) and 
then a letter representing the two week period of the year in which 
the discovery observations were acquired 
(example: L3f occured in the second half of March 2003).   
Discovery observations occurred between March 2003 and July 2005 plus one
block of fields (L7a) observed in January 2007.  
The CFEPS presurvey block \citep{2006Icar..185..508J} 
in 2002 also consisted of a single contiguous sky patch.
To enhance our sensitivity to the latitude distribution of the Kuiper belt we
also acquired two survey blocks of 11 square degrees each, at $\sim$10$^\circ$
ecliptic latitude (L5r) and $\sim$20$^\circ$ ecliptic latitude (L5s).  
Each of the discovery blocks was searched for TNOs
using our Moving Object Pipeline \citep[MOP; see][]{2004MNRAS.347..471P}.
Table~\ref{tab:fields} provides a summary of the survey fields, imaging
circumstances and detection thresholds, both for CFEPS and for the presurvey.
Figure~\ref{fig:cfeps_geom} presents
the sky coverage of our discovery blocks.  For a detailed description of the
initial CFEPS observing plan, field sequencing and follow-up strategy see \citet{2006Icar..185..508J} and P1.

\begin{deluxetable}{lrrrrrrlrccc}%[h]
\tabletypesize{\scriptsize}
\tablecolumns{12}
\tablewidth{0pt}
\tablecaption{Summary of Field positions and Detections.\label{tab:fields}}
\tablehead{
\colhead{Block}  &  \colhead{RA$^a$} &   \colhead{DEC$^a$}   &
\colhead{Fill$^b$} & \multicolumn{2}{c}{Charact. Det.$^c$} &
\colhead{Geometry} & \multicolumn{2}{c}{Discovery} & \colhead{limit$^d$} &
\multicolumn{2}{c}{Detection limits$^e$}\\ 
\colhead{\ } & \colhead{HRS} & \colhead{DEG} &  
\colhead{Factor} & \colhead{Disc.} & \colhead{Track.}  & \colhead{DEG x DEG} &
\colhead{date} & \colhead{filter} & \colhead{$g_{AB}$} &
\colhead{rate (``/h)} & \colhead{direction (DEG)}
}
\startdata
L3f   & 12:42  &  $-$04:33  &  0.80 &   3 &   2 &  4x4 & 2003-03-24  &  G.MP9401
& 23.75 & 1.7 to 5.1 & $-$10.0 to  50.0\\
L3h   & 13:03  &  $-$06:48  &  0.81 &  14 &  11 &  4x4 & 2003-04-26  &  R.MP9601
& 24.43$\dagger$ & 0.8 to 6.2 &   5.6 to  41.6\\
L3q   & 22:01  &  $-$12:04  &  0.89 &   9 &   7 &  4x4 & 2003-08-31  &  G.MP9401
& 24.08 & 1.2 to 6.2 & $-$38.0 to  $-$2.0\\
L3s   & 19:43  &  $-$01:20  &  0.87 &   5 &   5 & 14x1 & 2003-09-23  &  G.MP9401
& 23.95 & 0.8 to 8.0 & $-$42.6 to  $-$5.0\\
L3w   & 04:33  &   22:21  &  0.87 &  13 &  11 & 16x1 & 2003-12-16  &  G.MP9401
& 24.25 & 0.8 to 6.0 & $-$29.0 to  11.0\\ 
L3y   & 07:30  &   21:48  &  0.85 &  10 &  10 &  4x4 & 2003-12-24  &  G.MP9401
& 24.08 & 1.7 to 5.1 &  $-$6.0 to  24.0\\
\\
& & Total &  &  54 & 46 & \multicolumn{1}{c}{94 sqr. deg.}  \\
\\
L4h   & 13:35  &  $-$09:00  &  0.89 &  20 &  16 &  7x2; 1x1 & 2004-04-26  &  G.MP9401
& 24.06 & 0.8 to 6.0 &   2.0 to  42.0\\
L4j   & 15:12  &  $-$16:51  &  0.89 &  10 &  10 &  8x2 & 2004-04-25  &  G.MP9401
& 24.00 & 0.8 to 5.6 &  $-$3.6 to  36.4\\
L4k   & 15:12  &  $-$18:47  &  0.90 &  19 &  16 &  8x2 & 2004-05-24  &  G.MP9401
& 24.35 & 0.8 to 5.7 &  $-$1.0 to  35.0\\
L4m   & 19:14  &  $-$22:47  &  0.89 &   4 &   4 & 12x1 & 2004-06-25  &  G.MP9401
& 23.76 & 0.8 to 5.6 & $-$25.0 to  15.0\\
L4n   & 19:23  &  $-$21:33  &  0.90 &   4 &   4 & 14x1 & 2004-07-22  &  G.MP9401
& 23.74 & 0.8 to 6.0 & $-$27.7 to  12.3\\
L4o   & 19:15  &  $-$23:46  &  0.90 &   2 &   1 & 13x1 & 2004-07-24  &  G.MP9401
& 23.53 & 0.8 to 6.0 & $-$24.7 to  11.3\\
L4p   & 20:53  &  $-$18:27  &  0.85 &   9 &   9 &  8x2 & 2004-08-15  &  G.MP9401
& 24.00 & 1.0 to 5.7 & $-$30.0 to   0.0\\
L4q   & 21:26  &  $-$16:05  &  0.85 &  14 &  10 &  8x2 & 2004-08-19  &  G.MP9401
& 24.21 & 1.2 to 6.1 & $-$35.5 to  $-$0.5\\
L4v   & 02:35  &   15:10  &  0.78 &  18 &  14 & 2x2; 1x1; 5x2 & 2004-11-09  &  G.MP9401
& 24.40 & 0.8 to 6.3 & $-$34.0 to  $-$2.0\\
\\
& & Total &  & 100 & 84 & \multicolumn{1}{c}{133 sqr. deg.}  \\
\\
L5c   & 09:11  &   17:13  &  0.84 &  21 &  19 &  7x2; 1x1 & 2005-02-10  &  G.MP9401
& 24.30 & 0.8 to 6.4 &  $-$1.0 to  31.0\\
L5i   & 16:18  &  $-$22:18  &  0.90 &   7 &   7 &  8x2 & 2005-05-12  &  G.MP9401
& 23.84 & 0.4 to 7.3 &  $-$9.4 to  32.2\\
L5j   & 16:09  &  $-$19:59  &  0.89 &   3 &   3 &  8x2 & 2005-06-10  &  G.MP9401
& 23.49 & 0.4 to 7.0 &  $-$9.9 to  33.9\\
L5r   & 22:36  &   03:55  &  0.90 &   1 &   1 & 3x2; 1x1; 2x2 & 2005-09-03  &  G.MP9401
& 23.89 & 0.7 to 7.5 & $-$42.1 to  $-$1.9\\
L5s   & 22:28  &   14:35  &  0.90 &   1 &   1 & 3x2; 1x1; 2x2 & 2005-09-03  &  G.MP9401
& 24.00 & 0.7 to 7.5 & $-$41.8 to  $-$2.0\\
L7a   & 08:43  &   18:30  &  0.89 &  9 &  8 & patchy & 2007-01-19  &  G.MP9401
& 23.98 & 0.8 to 7.7 &  $-$4.1 to  34.9\\
\\
& & Total &  &  42 & 39 & \multicolumn{1}{c}{94 sqr. deg.}  \\
\\
& \multicolumn{2}{r}{Grand Total} &  &  196 & 169 & \multicolumn{1}{c}{321 sqr. deg.}  \\
\\
Pre   & 22:00  &  -13:00  &  0.90 &  13 &  10 &  3.5x2 & 2002-08-05  &  R &
24.85$^\star$ & 0.8 to 8.0 & $-$35.0 to $-$5.0\\
\\
\enddata
\tablecomments{($a$) RA/DEC is the approximate center of the
  field. ($b$) Fill Factor is the fraction of the rectangle
  covered by the mosaic and useful for TNO searching.
  ($c$) The number of objects in columns 5 and 6 correspond to those detected
  and tracked in the characterized sample, as defined in Sect.~\ref{sec:char}.
  ($d$) The limiting magnitude of the survey, $g_{AB}$, is in the SDSS
  photometric system and corresponding to a 40\% efficiency of detection. 
  ($e$)Detection limits give the limits on the sky motion in rate (``/hr) and
  direction (``zero degrees'' is due West, and positive to the North).
  $\dagger$ Although the L3h block was acquired in $\mpr$ filter, the reported
  limiting magnitude has been translated to $\mpg$ band by applying an offset
  of $\mpg - \mpr = 0.7$, which is the average $\mpg - \mpr$ color of our full
  sample (see Table~\ref{tab:phot}).
  $\star$ The Presurvey block was acquired in $R$ filter with the CFH12K camera
  \citep{2006Icar..185..508J}. The limiting magnitude has been translated to
  $\mpg$ band by applying  an offset of $\mpg - R = 0.8$.
}
\end{deluxetable}

\section{Sample Characterization}
\label{sec:char}

The photometric calibration of the discovery triplets to a common reference
frame and determination of our detection efficiency is required for our survey
simulator analysis.  It is presented in Appendix~\ref{sec:app_b} and the
photometric measurements of all CFEPS TNOs
acquired in photometric conditions are given in Table~\ref{tab:phot}.  

We characterized the magnitude-dependent detection probability of each
discovery block by inserting artificial sources in the images and running these
images through our detection pipeline to recover these artificial sources.
We used the $\mpg$\ filter at CFHT for all our discovery observations, except for
block L3h which was acquired using the $\mpr$\ filter. For that block's fields,
we shifted the limits to a nominal $\mpg$\ value using a color of
($\mpg-\mpr$) = 0.70, corresponding to the mean ($\mpg-\mpr$) color of our full
CFEPS sample.  The TNOs in each block that have a magnitude brighter
than that block's 40\% detection probability are considered to be part of the
CFEPS {\em characterized sample}.
Because detection efficiencies below $\sim$ 40\% determined by human operators
and our MOP diverge - MOP accepts more faint objects, at the expense of false
detections - \citep{2004MNRAS.347..471P}, and since characterization is
critical to the CFEPS goals, we chose not to utilize the sample faint-ward of
the measured 40\% detection-efficiency level for quantitative science (although
we report these discoveries, many of which were tracked to precise orbits).
The {\it characterized} CFEPS sample consists of \NKBO\ objects of the
\lKBO\ discovered (see Table~\ref{tab:phot} for a list of these TNOs).
The fraction of objects detected bright-ward of our cutoff is 
consistent with the shape of the TNO luminosity function
\citep{2008ssbn.book...71P} and typical decay in detection efficiency
due to gradually increasing stellar confusion and the rapid fall-off at
the SNR limit.

Our discovery and tracking observations were made using short exposures
designed to maximize the efficiency of detection and tracking of the TNOs in
the field.
These observations do not provide the high-precision flux measurements
necessary for possible classification based on broadband colors of TNOs and we
do not comment here on this aspect of the CFEPS sample.

\section{Tracking and Lost Objects}

Tracking during the first opposition was done using the built-in followup of
the CFEPS project.
Subsequent tracking, over the next 3 oppositions, occurred at a
variety of facilities, including CFHT.  The observational efforts outside CFHT
are summarized in Table~\ref{tab:followup}.  In spring 2006 the CFEPS project
made an initial data release of the complete observing record for the L3
objects (objects discovered in 2003; before all the refinement observations for
all objects were complete).  The L3 release was reported to the Minor Planet
Center (MPC)
\citep{2006MPEC....H...29G,2006MPEC....H...35K,2006MPEC....H...30K} and
additional followup that has occurred since the 2006 release has also been
reported to the MPC.  The final release of the complete observing record for
all remaining CFEPS objects is available from the MPC
\citep{2011MPEC...........}.  Detailed astrometric and photometric data for the
CFEPS objects can be found on the CFEPS specific databases\footnote{{\it
    http://www.cfeps.net/tnodb/}, {\it http://www.obs-besancon.fr/bdp/}}.
The correspondence between CFEPS internal designations and MPC designations can
be determined using Tables~\ref{tab:chclass} and \ref{tab:nonchclass}, 
or from electronic tables on the {\it cfeps.net} site.  
All
characterized and tracked objects are prefixed by {\it L} and are used with
the survey simulator for our modeling below.  The tracking observations provide
sufficient information to allow reliable orbits to be determined such that
unambiguous dynamical classification can be achieved in nearly all cases.
Ephemeris errors are smaller than a few tens of arc-seconds over the next 5
years.  Our standard was to pursue tracking observations until the semimajor
axis uncertainty was $<0.1$\%; in Tables~\ref{tab:chclass} and
\ref{tab:nonchclass}, orbital elements are shown to the precision with which
they are known, with typical fractional accuracies on the order of $10^{-4}$ or
better.  In the cases of resonant objects even this precision may not be enough
to determine the amplitude of the resonant argument.

\begin{deluxetable}{llr}
\tablecolumns{3}
\tablewidth{0pc}
\tablecaption{Follow-up/Tracking Observations.\label{tab:followup}}
\tablehead{
\colhead{UT Date}    & \colhead{Telescope} & \colhead{No. Obs.}} 
\startdata

2002 Aug 05 & CFHT + 12k            &   6 \\
2002 Sep 03 & NOT 2.56m             &   6 \\
2002 Sep 02 & Calar-Alto 2.2-m      &   9 \\
2002 Sep 30 & CFHT 3.5-m            &   6 \\
2002 Nov 28 & CFHT 3.5-m            &  10 \\
2003 Jul 26 & ESO 2.2m              &   6 \\
2004 Feb 19 & WIYN 3.5-m            &   4 \\
2004 Apr 15 & Hale 5-m    &  73 \\
2004 May 24 & Mayall 3.8-m          &   6 \\
2004 Aug 12 & CFHT 3.5m             &  15 \\
2004 Sep 06 & KPNO 2m               &  15 \\
2004 Sep 11 & Mayall 3.8-m          &  25 \\
2004 Sep 16 & Hale 5-m              &  20 \\
2004 Sep 21 & CFHT 3.5m             &   4 \\
2005 Jul 08 & Gemini-North 8-m      &  45 \\
2005 Jul 09 & Hale 5-m              &  47 \\
2005 Jul 11 & ESO 2.2m              &  25 \\
2005 Aug 01 & VLT UT-1              &  53 \\
2005 Sep 24 & WIYN 3.5-m            &   9 \\
2005 Oct 03 & Hale 5-m              &  72 \\
2005 Nov 04 & Mayall 3.8m           &  31 \\
2005 Dec 04 & MDM 2.4-m             &  10 \\
2006 Jan 28 & Hale 5-m              &  50 \\
2006 May 01 & CFHT 3.5m             &  23 \\
2006 May 02 & WIYN 3.5-m            &  32 \\
2006 May 26 & CFHT 3.5m             &  20 \\
2006 Jun 25 & Mayall 3.8-m          &   2 \\
2006 Jul 03 & CFHT 3.5m             &  18 \\
2006 Jul 26 & Hale 5-m              &  15 \\
2006 Sep 18 & CFHT 3.5m             &   7 \\
2006 Sep 26 & MMT 6.5m              &  11 \\
2006 Oct 22 & Hale 5-m              &  29 \\
2006 Oct 21 & WHT 4m                &  17 \\
2006 Nov 23 & WIYN 3.5-m            &  41 \\
2007 Feb 14 & 2.1-m reflector       &   3 \\
2007 Feb 21 & Hale 5-m              &  22 \\
2007 May 15 & Hale 5-m              &  45 \\
2007 May 15 & KPNO 2m               &  23 \\
2007 Jun 02 & MMT 6.5m              &   3 \\
2007 Sep 11 & WIYN 3.5-m            &  32 \\
2007 Sep 16 & Hale 5-m              &  27 \\
2007 Nov 08 & WIYN 3.5-m  &  30 \\
2008 May 03 & WIYN 3.5-m  &  52 \\
2008 Jun 07 & CTIO 4-m              &  28 \\
2008 Oct 23 & WIYN 3.5-m  &   3 \\
2008 Dec 06 & Hale 5-m              &   9 \\
2009 Jan 26 & CFHT 3.5m             &  19 \\
2009 Apr 17 & MMT 6.5m              &   3 \\
2009 Apr 23 & Subaru 8-m            &   1 \\
2009 Jun 20 & WIYN 3.5-m  &  22 \\

\enddata
\tablecomments{
UT Date is the start of the observing run; No. Obs. is the number of
astrometric measures reported from the observing run. Only observations
not part of the Very Wide component of CFHT-LS are reported here.
Runs with low numbers of astrometric measures were either wiped out by poor
weather, or not meant for CFEPS objects follow-up originally.
}
\end{deluxetable}

Of the \NKBO TNOs in our CFEPS characterized sample \TKBO have been tracked
through 3 oppositions or more (ie. not lost) and their orbits are now known to
a precision of ${\Delta a}/{a} < 0.1 \%$ and can be reliably classified into
orbital sub-populations (see below).  The very high fraction of our
characterized sample for which classification is possible (86\%) is by far the
largest `tracking fraction' among large scale TNO surveys to-date and is 
due to the strong emphasis on followup observations in our observing strategy,
made possible thanks to the time allocation committees of the many
observatories listed in Table~\ref{tab:followup}.

The initial tracking of TNOs discovered by CFEPS is through blind return to the
discovery fields to ensure that there is no orbital bias in the tracked
fraction. We do find, however, that the tracked fraction is a function of the
magnitude of the TNO and have characterized this bias. For the full CFEPS
fields we find the same magnitude dependance as for the L3 fields for objects
brighter than the limit of the characterized sample, which we model as

$$
f_{t,L7}(\mpg) = 
\cases{ 1.0 & ($\mpg \leqslant 22.8$) \cr
        1.0 -0.25(\mpg - 22.8) & ($\mpg > 22.8$) 
      }
$$

\noindent where $f_{t,L7}$ is the tracked fraction.  
The tracked fraction remains well above 50\%
down to the characterized limit of the survey blocks.
We have also re-examined the magnitude dependence of the tracked fraction 
of our Pre-survey discoveries \citep{2006Icar..185..508J} and find 
$$
f_{t,L7}(\mpg) = 
\cases{ 1.0 & ($\mpg \leqslant  24.1$) \cr
        1.0 - 2.5(\mpg - 24.1) & ($\mpg > 24.1$) 
      }.
$$
The Pre-survey observations used much longer exposure times than for CFEPS,
hence the deeper limiting magnitude reached.
We also had a smaller survey area and were able to perform a more thorough
follow-up campaign, resulting in a tracking efficiency that essentially was
100\% up to the limiting magnitude of the discoveries.
Our Pre-survey discovery observations where reported on the Landolt-R system
and we have transformed our Pre-survey limits to $\mpg$, for use in our
survey simulator, using a constant color offset of ($\mpg$ - R) = 0.8
\citep{2002A&A...389..641H}.

\subsection{Orbit Classification}

We adopt the convention that, based on orbital elements and dynamical behavior,
the Kuiper Belt can be divided into three broad orbital classes. An object is
checked against each dynamical class in the order below to decide whether or
not it belongs to that class, each object can belong to only one class.  A schematic representation of this dynamical classifaction is shown in Figure~1
of \citet{2008ssbn.book...43G}.

\begin{itemize}
\item resonant (objects currently in a mean-motion resonance with
  Neptune) 
\item scattering (objects that over 10~Myr forward in time integrations
  experience encounters with Neptune resulting in variation of semimajor-axis
  of more than 1.5~AU) 
\item classical or detached belt (everything that remains).
  One further sub-divides the classical belt into:
\begin{itemize} 
\item inner classical belt (objects with semi-major axis interior to
  the 3:2 MMR)
\item main classical belt (objects whose semi-major axis is between the 3:2 and
  2:1 MMRs)
\item outer classical belt (objects with semi-major axis exterior to the 2:1
  MMR with $e<0.24$)
\item detached (those objects with semi-major axis beyond the 2:1 MMR
  that have $e>0.24$)
\end{itemize}
\end{itemize}

The classical belt is often also divided into high-inclination and 
low-inclination
objects.
For the L7 model, we work from the hypothesis described in \citet{2001AJ....121.2804B} that there exist two distinct populations, one with a
wide inclination distribution (the `hot' population), and the other one with a
narrow inclination distribution (the `cold' population), with both populations overlapping with each other in 
inclination space (thus some ``cold'' objects may have large inclination, 
and some ``hot'' objects may have low inclination).
In the literature, the separation between hot and cold populations is sometimes presented as
a sharp cut in inclination, often around 5$^\circ$, under the assumption that
an object with inclination less (greater) than that threshold has a very
high likelihood to be a member of the cold (hot) population.
As will be seen in Section~\ref{sec:lumfun}, the veracity of this assumption depends on the
physical size of the objects being sorted, larger objects ($H<$7) having a much higher probability of being
from the hot population, regardless of their inclination, while the objects from the cold
population dominate at smaller (H$>$8) sizes.
A strict inclination cut does not isolate the two mixed populations.

Following the procedure in \citet{2008ssbn.book...43G} 
\cite[similar to][]{2003AJ....126..430C}, 
we extend the L3 sample classification given in P1 to our full CFEPS sample 
as of November 2009 (including all refinement observations to that date).
Using this classification procedure, 15 of our objects remain 
insecure (even though these have observational arcs extending across 5 
oppositions!); all of these are due to their proximity to a resonance 
border where the remaining astrometric uncertainty makes it unclear if 
the object is actually resonant. 
We list these `insecure' objects in the category shown by 2 of the 3 clones.
Table~\ref{tab:chclass} gives the classification of all characterized objects
used for comparison with the Survey Simulator's artificial detections.
Several objects had been independently discovered before we submitted our
observations to the MPC and are marked with a PD suffix.
Although we do not claim `discoverer credit' for these objects, 
they have just as much scientficially-exploitable value because
they were detected during our characterized observations and hence can to be
included when running our survey simulator.
Table~\ref{tab:nonchclass} gives the classification of the tracked 
objects below the 40\% efficiency threshold, hence deemed 
non-characterized and not used in our Survey Simulator comparisons.

\ 

\begin{deluxetable}{ll|llrrl}
\tabletypesize{\scriptsize}
\tablecolumns{7}
\tablewidth{0pc}
\tablecaption{Characterized Object Classification.\label{tab:chclass}}
\tablehead{
\multicolumn{2}{c|}{DESIGNATIONS} & \colhead{a} & \colhead{e} &
\colhead{i} & \colhead{dist} & \colhead{Comment}\\
\colhead{CFEPS} & \multicolumn{1}{l|}{MPC} & \colhead{AU} & \colhead{\ }  &
\colhead{$^\circ$} &  \colhead{AU}  & \colhead{\ }
}
\startdata

\cutinhead{Resonant Objects}

%\loaddata{tables/ResonantTableCh}
L3y11   & (131697) 2001 XH255 & 34.925 & 0.0736  &  2.856 & 34.0  &  5:4 \quad MPC$_W$   \\
\\
  L4h14 & 2004 HM79  &  36.441 & 0.07943 &  1.172 & 38.0 &  4:3 \\ 
L3s06   & (143685) 2003 SS317 & 36.456 & 0.2360  &  5.905 & 28.2  &  4:3 \\
  L5c23 & 2005 CF81  &  36.473 & 0.06353 &  0.405 & 34.4 &  4:3 \\ 
  L7a10 & 2005 GH228 &  36.663 & 0.18814 & 17.151 & 30.6 &  4:3 \quad I \\ 
\\
  L4k11 & 2004 KC19  &  39.258 & 0.23605 &  5.637 & 30.2 &  3:2  \\ 
  L4h15 & 2004 HB79  &  39.260 & 0.22862 &  2.661 & 32.0 &  3:2  \\ 
  L5c11 & 2005 CD81  &  39.262 & 0.15158 & 21.344 & 45.2 &  3:2  \\ 
  L4h06 & 2004 HY78  &  39.302 & 0.19571 & 12.584 & 31.8 &  3:2  \\ 
  L4v18 & 2004 VY130 &  39.342 & 0.27616 & 10.203 & 28.5 &  3:2  \\ 
  L4m02 & 2004 MS8   &  39.344 & 0.29677 & 12.249 & 27.8 &  3:2  \\ 
L3s02   & 2003 SO317 & 39.346 & 0.2750  &  6.563 & 32.3  &  3:2  \\
L4h09PD & (47932) 2000 GN171 &  39.352 & 0.28120 & 10.815 & 28.5 &  3:2  \\ 
L3h19   & 2003 HF57  & 39.36  & 0.194   &  1.423 & 32.4  &  3:2   \\
L3w07   & 2003 TH58  & 39.36  & 0.0911  & 27.935 & 35.8  &  3:2   \\
  L4h07 & 2004 HA79  &  39.378 & 0.24697 & 22.700 & 38.4 &  3:2  \\ 
L3h11   & 2003 HA57  & 39.399 & 0.1710  & 27.626 & 32.7  &  3:2   \\
L3w01   & 2005 TV189 & 39.41  & 0.1884  & 34.390 & 32.0  &  3:2   \\
  L4j11 & 2004 HX78  &  39.420 & 0.15270 & 16.272 & 33.6 &  3:2  \\ 
  L4v09 & 2004 VX130 &  39.430 & 0.20696 &  5.745 & 34.8 &  3:2  \\ 
L3h14   & 2003 HD57  & 39.44  & 0.179   &  5.621 & 32.9  &  3:2   \\
L3s05   & 2003 SR317 & 39.44  & 0.1667  &  8.348 & 35.5  &  3:2   \\
  L4v13 & 2004 VV130 &  39.454 & 0.18827 & 23.924 & 32.8 &  3:2  \\ 
  L4k01 & 2004 KB19  &  39.484 & 0.21859 & 17.156 & 39.5 &  3:2  \\ 
L3h01   & 2004 FW164 & 39.492 & 0.1575  &  9.114 & 33.3  &  3:2   \\
L5i06PD & 2001 KQ77  &  39.505 & 0.15619 & 15.617 & 36.2 &  3:2  \\ 
L4h10PD & 1995 HM5   &  39.521 & 0.25197 &  4.814 & 31.1 &  3:2  \\ 
  L4v12 & 2004 VZ130 &  39.551 & 0.28159 & 11.581 & 29.2 &  3:2  \\ 
  L4h08 & 2004 HZ78  &  39.580 & 0.15095 & 13.310 & 34.8 &  3:2  \\ 
\\
  L5c08 & 2006 CJ69  &  42.183 & 0.22866 & 17.916 & 35.5 &  5:3  \\ 
L3y06   & 2003 YW179 & 42.193 & 0.1537  &  2.384 & 35.7  &  5:3   \\
L5c13PD & 1999 CX131 &  42.240 & 0.23387 &  9.757 & 41.8 &  5:3  \\ 
  L4v05 & 2004 VE131 &  42.297 & 0.25889 &  5.198 & 39.6 &  5:3  \\ 
L3y12PD & (126154) 2001 YH140 & 42.332 & 0.14043 & 11.078 & 36.4 &  5:3  \\
  L4k10 & 2004 KK19  &  42.410 & 0.14391 &  4.485 & 46.0 &  5:3 \quad I \\ 
\\
L3q08PD & (135742) 2002 PB171 & 43.63  & 0.125   &  5.450 & 40.7  &  7:4  \\ 
  L4n03 & 2004 OQ15  &  43.646 & 0.12472 &  9.727 & 40.5 &  7:4  \\ 
L3w03   & 2003 YJ179 &  43.66  & 0.0794  &  1.446 & 40.3 &  7:4  \\
  L4v10 & 2004 VF131 &  43.672 & 0.21492 &  0.816 & 42.0 &  7:4  \\ 
K02O03  & 2000 OP67  &  43.72  & 0.19  1 &  0.751 & 39.3 &  7:4 \\
\\
  L4h11 & 2004 HN79  &  45.736 & 0.22936 & 11.669 & 37.4 & 15:8 \quad I \\ 
\\
  L4h18 & 2004 HP79  &  47.567 & 0.18250 &  2.253 & 39.5 &  2:1  \\ 
  L4k16 & 2004 KL19  &  47.660 & 0.32262 &  5.732 & 32.3 &  2:1  \\ 
  L4k20 & 2004 KM19  &  47.720 & 0.29180 &  1.686 & 33.8 &  2:1  \\ 
K02O12  & 2002 PU170 & 47.75   & 0.2213  &  1.918 & 47.2 &  2:1  \\
  L4v06 & 2004 VK78  &  47.764 & 0.33029 &  1.467 & 32.5 &  2:1  \\ 
\\
L3y07   & (131696) 2001 XT254 & 52.92  & 0.3221  &  0.518 & 36.6  &  7:3 \quad  MPC$_W$   \\
L5c19PD & 2002 CZ248 &  53.039 & 0.38913 &  5.466 & 36.2 &  7:3  \\ 
\\
  L5c12 & 2002 CY224 &  53.892 & 0.34651 & 15.733 & 36.3 & 12:5  \\ 
\\
  L4j08 & 2004 HO79  &  55.206 & 0.41166 &  5.624 & 37.3 &  5:2  \\ 
L3f04PD & (60621) 2000 FE8 & 55.29  & 0.4020  &  5.869 & 36.0  &  5:2  \\ 
L4j06PD & 2002 GP32  &  55.387 & 0.42195 &  1.559 & 32.1 &  5:2  \\ 
  L4k14 & 2004 KZ18  &  55.419 & 0.38191 & 22.645 & 34.4 &  5:2  \\ 
L4h02PD & 2004 EG96  &  55.550 & 0.42291 & 16.213 & 32.2 &  5:2  \\ 
\\
  L4v08 & 2004 VU130 &  62.194 & 0.42806 &  8.024 & 49.7 &  3:1  \\ 
\\
L3y02   & 2003 YQ179 & 88.38  & 0.5785  & 20.873 & 39.3  &  5:1 \quad I \\

\cutinhead{Inner Classical Belt}  

%\loaddata{tables/InnerBeltCh}
L3y14PD & (131695) 2001 XS254 &  37.220 & 0.05211 &  4.262 & 35.3 & I (11:8) \\
L4q12PD & 2000 OB51  &  37.820 & 0.03501 &  4.458 & 36.6 & \\ 
  L4q10 & 1999 OJ4   &  38.017 & 0.02539 &  4.000 & 38.1 & \\ 
  L4k18 & 2004 KD19  &  38.257 & 0.01707 &  2.126 & 38.9 & \\ 
  L4o01 & 2004 OP15  &  38.584 & 0.05532 & 22.946 & 38.7 & \\ 
L3w06   & 2003 YL179 &  38.82  & 0.002   &  2.525 & 38.7 & \\

\cutinhead{Main Classical Belt} 

%\loaddata{tables/ClassicalBeltCh}
  L4k12 & 2004 KH19  &  40.772 & 0.11721 & 35.230 & 43.6 & \\ 
  L4q05 & 2004 QE29  &  40.878 & 0.08372 & 24.125 & 37.5 & \\ 
  L4k19 & 2005 JB186 &  41.471 & 0.10588 & 20.220 & 38.0 & \\ 
L3w05   & 2003 YK179 &  41.67  & 0.146   & 19.605 & 42.7 & \\
  L4h16 & 2004 HL79  &  42.126 & 0.07520 & 16.759 & 40.0 & \\ 
L5s01PD & (120347) 2004 SB60  &  42.028 & 0.10667 & 23.931 & 43.7 & \\ 
L3s01   & 2003 SN317 &  42.50  & 0.0421  &  1.497 & 41.5 & \\
  L4q15 & 1999 ON4   &  42.571 & 0.03995 &  3.187 & 40.9 & \\ 
L3h05   & 2003 HY56  &  42.604 & 0.037   &  2.578 & 42.5 & \\
L3q02PD & 2001 QB298 &  42.618 & 0.0962  &  1.800 & 39.1 & \\
L3s03   & 2003 SQ317 &  42.63  & 0.0795  & 28.568 & 39.3 & \\
K02O20  & 2002 PV170 &  42.643 & 0.016   & 1.271 & 42.2  & \\
K02P32  & 2002 PX170 &  42.65  & 0.041   & 1.570 & 42.8  & \\
  L5c03 & 2005 CE81  &  42.715 & 0.04666 &  3.084 & 40.8 & \\ 
  L5i01 & 2006 HA123 &  42.778 & 0.04615 &  3.303 & 41.0 & \\ 
  L4p02 & 2004 PU117 &  42.817 & 0.01461 &  1.874 & 42.4 & \\ 
  L4p01 & 2004 PT117 &  42.983 & 0.04115 &  1.238 & 43.6 & \\ 
K02O40  & 2002 PY170 &  43.015 & 0.030   &  3.016 & 43.0 & \\
  L4q03 & 2004 QD29  &  43.020 & 0.11388 & 23.862 & 40.6 & I (12:7) \\ 
L3w11   & 2003 TK58  &  43.078 & 0.0647  &  3.355 & 45.6 & \\
  L4m03 & 2004 MT8   &  43.120 & 0.04195 &  2.239 & 44.9 & \\ %Faint \\ 
L4h05PD & 2001 FK185 &  43.255 & 0.03994 &  1.171 & 41.7 & \\ 
  L4j10 & 2004 HH79  &  43.259 & 0.06010 &  8.610 & 43.2 & \\ 
  L4j02 & 2004 HF79  &  43.269 & 0.02547 &  1.484 & 42.4 & \\ 
  L4k04 & 2004 KG19  &  43.272 & 0.02164 &  0.963 & 42.4 & \\ 
L5c07PD & 2005 XU100 &  43.398 & 0.10283 &  7.869 & 41.7 & \\ 
  L7a06 & 2006 WF206 &  43.500 & 0.04246 &  2.056 & 44.4 & \\ 
L3w10   & 2003 TL58  &  43.542 & 0.0456  &  7.738 & 42.2 & \\
L3y01   & 2003 YX179 &  43.582 & 0.044   &  4.850 & 42.5 & \\
L3y05   & 2003 YS179 &  43.585 & 0.022   &  3.727 & 43.8 & \\
L3h18   & 2003 HG57  &  43.612 & 0.0323  &  2.098 & 43.0 & \\
  L4p05 & 2004 PW117 &  43.620 & 0.06023 &  1.862 & 46.0 & \\
  L7a05 & 2005 BV49  &  43.684 & 0.04575 &  7.981 & 41.8 & \\ 
  L5j04 & 2005 LB54  &  43.690 & 0.04752 &  3.006 & 41.8 & \\ 
L4h01PD & (181708) 1993 FW    &  43.717 & 0.04807 &  7.750 & 41.9 & \\
L4p06PD & 2001 QY297 &  43.835 & 0.08332 &  1.547 & 42.8 & \\ 
  L4h12 & 2004 HK79  &  43.888 & 0.07800 &  1.946 & 41.3 & \\ 
L5i03PD & 2001 KO77  &  43.898 & 0.14569 & 20.726 & 37.7 & \\ 
  L4h13 & 2004 HJ79  &  43.947 & 0.04419 &  3.317 & 45.0 & \\ 
  L4v03 & 2004 VC131 &  43.951 & 0.07395 &  0.490 & 40.7 & \\ 
  L5c22 & 2007 DS101 &  43.991 & 0.08474 &  1.389 & 44.6 & \\ 
L3h13   & 2003 HH57  &  44.04  & 0.088   &  1.436 & 40.2 & \\
L3h09   & 2003 HC57  &  44.05  & 0.072   &  1.038 & 43.4 & \\
  L5i05 & 2005 JY185 &  44.077 & 0.06848 &  2.139 & 44.6 & \\ 
  L7a07 & 2005 BW49  &  44.097 & 0.07959 &  2.102 & 41.9 & \\ 
L3q06PD & 2001 QJ298 &  44.10  & 0.0388  &  2.151 & 45.2 & \\
  L4k03 & 2004 KF19  &  44.123 & 0.06348 &  0.108 & 41.4 & \\ 
L5c21PD & 2005 EE296 &  44.126 & 0.06804 &  3.296 & 46.2 & \\ 
L3q09PD & 2001 QX297 &  44.15  & 0.0275  &  0.911 & 43.5 & \\
  L5c18 & 2007 CS79  &  44.159 & 0.03582 &  1.540 & 42.8 & \\ 
L3h20   & 2003 HE57  &  44.17  & 0.100   &  8.863 & 40.0 & \\
L5c24PD & 1999 CU153 &  44.172 & 0.06520 &  2.698 & 42.7 & \\ 
  L4v02 & 2004 VB131 &  44.189 & 0.07267 &  1.747 & 46.5 & \\ 
  L4j03 & 2004 HG79  &  44.200 & 0.02298 &  3.595 & 43.2 & \\ 
  L4p09 & 2004 PX117 &  44.261 & 0.09965 &  3.747 & 46.1 & \\ 
L4p08PD & 2001 QZ297 &  44.283 & 0.06442 &  1.856 & 42.0 & \\ 
  L5j03 & 2005 LA54  &  44.314 & 0.06719 &  7.919 & 41.6 & \\ 
  L4j01 & 2004 HE79  &  44.316 & 0.09805 &  3.089 & 40.0 & \\ 
K02P41  & 2002 PA171 &  44.34  & 0.076   & 2.511 & 47.7  & \\
  L4k02 & 2004 KE19  &  44.360 & 0.04981 &  1.178 & 42.6 & \\ 
L3w08   & 2003 TJ58  &  44.40  & 0.0864  &  0.954 & 40.8 & \\
L3w02   & 2003 TG58  &  44.54  & 0.103   &  1.660 & 43.7 & I (9:5) \\
L3w04   & (143991) 2003 YO179 &  44.602 & 0.1370  & 19.393 & 41.3 & \\
  L5i08 & 2005 JJ186 &  44.636 & 0.09431 &  4.141 & 41.8 & \\ 
K02O43  & 2002 PC171 &  44.706 & 0.059   & 3.574 & 42.7  & \\
  L4n04 & 2004 MU8   &  44.856 & 0.08180 &  3.580 & 48.2 & \\ 
K02O32  & 2002 PW170 &  44.88  & 0.074   & 3.933 & 47.4  & \\
  L4q16 & (66452) 1999 OF4 &  44.933 & 0.06380 &  2.660 & 45.2 & \\ 
  L4j12 & 2006 JV58  &  44.961 & 0.06094 &  0.317 & 42.2 & \\ 
L5c20PD & 2002 CZ224 &  44.980 & 0.06304 &  1.687 & 47.7 & I (11:6) \\ 
L3w09   & 2004 XX190 &  45.171 & 0.1042  &  1.577 & 40.9 & \\
L5c10PD & 1999 CJ119 &  45.325 & 0.06651 &  3.205 & 42.3 & \\ 
  L5c06 & 2007 CQ79  &  45.441 & 0.07721 &  1.185 & 45.8 & \\ 
  L4v01 & 2004 VA131 &  45.538 & 0.09613 &  0.767 & 41.2 & \\ 
L4k15PD & 2003 LB7   &  45.580 & 0.13130 &  2.294 & 40.1 & \\ 
  L5c02 & 2006 CH69  &  45.735 & 0.03535 &  1.791 & 44.2 & \\ 
  L4q11 & 1999 OM4   &  45.924 & 0.11643 &  2.088 & 44.0 & \\ 
  L4j07 & 2004 HD79  &  45.941 & 0.03205 &  1.305 & 47.3 & \\ 
L5i02PD & 2001 KW76  &  46.013 & 0.21613 & 10.460 & 39.6 & \\ 
  L4p03 & 2004 PV117 &  46.069 & 0.15343 &  4.324 & 39.5 & \\ 
L7a04PD & 2002 CY248 &  46.191 & 0.14635 &  7.038 & 51.8 & \\ 
  L4k13 & 2006 JU58  &  46.239 & 0.12464 &  7.035 & 46.5 & \\ 
L3q04PD & 2002 PT170 &  46.24  & 0.143   &  3.703 & 50.5 & \\
  L4v14 & 2004 VD131 &  46.324 & 0.12253 &  3.646 & 41.5 & \\ 
  L4j05 & 2004 HC79  &  46.399 & 0.16064 &  1.446 & 39.0 & \\ 
  L4q09 & 2000 PD30  &  46.519 & 0.02232 &  4.594 & 45.7 & \\ 
L3y03   & 2003 YU179 &  46.75  & 0.1597  &  4.855 & 39.6 & \\
  L4k17 & 2004 KJ19  &  46.967 & 0.23543 & 24.421 & 38.5 & \\ 
L7a11PD & 2000 CO105 &  47.046 & 0.14750 & 19.270 & 49.3 & \\ 
L3y09   & 2003 YV179 &  47.10  & 0.222   & 15.569 & 41.1 & \\
  L5c14 & 2007 CR79  &  47.149 & 0.21876 & 21.869 & 36.9 & \\ 
L3h04   & 2003 HX56  &  47.196 & 0.2239  & 29.525 & 45.5 & \\
  L4m04 & 2004 MV8   &  47.234 & 0.17503 & 27.205 & 39.1 & \\

\cutinhead{Outer Classical Belt}

%\loaddata{tables/OuterBeltCh}
  L4q06 & 2004 QG29  &  48.480 & 0.23517 & 27.134 & 37.8 & \\ 
  L4q14 & 2004 QH29  &  50.859 & 0.22922 & 12.010 & 39.9 & \\ 
  L5c16 & 2005 CG81  &  53.834 & 0.23684 & 26.154 & 44.6 & \\

\cutinhead{Detached Classical Belt}

%\loaddata{tables/DetachedDiskCh}
  L5i04 & 2005 JK186 &  47.264 & 0.24363 & 27.252 & 38.1 & \\ 
L3q03   & 2003 QX113 & 49.55  & 0.252   & 6.753 & 58.3 & \\
  L7a02 & 2006 WG206 &  50.416 & 0.29111 & 14.297 & 38.9 & \\ 
L4p04PD & 2000 PE30  &  54.318 & 0.34216 & 18.416 & 37.6 & \\ 
  L5c15 & 2005 CH81  &  55.156 & 0.31812 &  5.136 & 37.6 & \\ 
  L4n06 & 2004 OS15  &  55.760 & 0.31667 &  4.248 & 39.5 & \\ 
  L4n05 & 2004 OR15  &  56.248 & 0.33882 &  6.919 & 37.3 & \\ 
L3f01   & 2003 FZ129 & 61.71  & 0.3840 &  5.793 & 38.0 & \\
  L4h21 & 2004 HQ79  &  63.299 & 0.42264 &  6.473 & 36.6 & \\ 
  L5j02 & 2005 LC54  &  67.354 & 0.46279 & 22.443 & 43.1 & I (10:3) \\ 
  L5r01 & 2005 RH52  & 153.800 & 0.74644 & 20.447 & 39.0 & I scattering \\

\cutinhead{Scattering Disk}

%\loaddata{tables/ScatteringDiskCh}
  L4k09 & 2004 KV18  &  30.192 & 0.18517 & 13.586 & 26.6 & \\ 
  L4m01 & 2004 MW8   &  33.479 & 0.33308 &  8.205 & 31.4 & \\ 
  L4p07 & 2004 PY117 &  39.953 & 0.28088 & 23.545 & 29.6 & \\ 
L3q01   & 2003 QW113 &  50.99 & 0.484  &  6.922 & 38.2 &    \\
  L7a03 & 2006 BS284 &  59.613 & 0.43949 &  4.575 & 47.0 & \\ 
  L4v11 & 2004 VH131 &  60.036 & 0.62928 & 11.972 & 26.8 & \\ 
  L4v04 & 2004 VG131 &  64.100 & 0.50638 & 13.642 & 31.8 & \\ 
L3h08   & 2003 HB57  & 159.6  & 0.7613 & 15.499 & 38.4 &    \\

\enddata
\tablecomments
{M:N: object in the M:N resonance;
 I: indicates that the orbit classification is insecure
 (see \citet{2008ssbn.book...43G} for an explanation of the exact meaning);
 (M:N): object may be in the M:N resonance;
 MPC$_W$: indicates object was in MPC database but found $+1^\circ$ from
 predicted location. Objects prefixed with {\it L} are the characterized,
 tracked objects discovered during CFEPS; objects prefixed with {\it K02} were
 discovered in our pre-survey \citep{2006Icar..185..508J};
 The full orbital elements are available in electronic form from either {\it
    http://www.cfeps.net/tnodb/} or the MPC.
}

\end{deluxetable}

\begin{deluxetable}{ll|llrrl}
\tabletypesize{\scriptsize}
\tablecolumns{7}
\tablewidth{0pc}
\tablecaption{Non Characterized Object Classification.\label{tab:nonchclass}}
\tablehead{
\multicolumn{2}{c|}{DESIGNATIONS} & \colhead{a} & \colhead{e} &
\colhead{i} & \colhead{dist} & \colhead{Comment}\\
\colhead{CFEPS} & \multicolumn{1}{l|}{MPC} & \colhead{AU} & \colhead{\ }  &
\colhead{$^\circ$} &  \colhead{AU}  & \colhead{\ }
}
\startdata

\cutinhead{Resonant Objects}

%\loaddata{tables/ResonantTableNonCh}
  U5j06 &   &  39.369 & 0.22055 & 13.525 & 31.2 &  3:2  \\ 
\\
U3s04   & 2003 SP317 & 45.961 & 0.1694  &  5.080 & 44.9  & 17:9  \quad I \\
\\
  U7a08 &   &  47.702 & 0.19600 &  7.020 & 38.4 &  2:1  \\ 
\\
U5j01PD & (136120) 2003 LG7 &  62.157 & 0.47825 & 20.104 & 33.1 &  3:1  \\

\cutinhead{Main Classical Belt} 

%\loaddata{tables/ClassicalBeltNonCh}
U3w13   & 2003 YM179 & 40.960 & 0.056  & 23.414 & 40.2  &   \\
U3f02   & 2003 FA130 & 42.602 & 0.031  &  0.288 & 41.3  &   \\
U4j09 &   &  42.642 & 0.00775 &  3.044 & 42.3 & \\ 
U7a09 &   &  42.701 & 0.09231 &  2.931 & 44.8 & \\ 
U3w17   & 2002 WL21 & 43.103 & 0.0415 &  2.552 & 41.6  &   \\
U3y16   & 2003 YR179 & 43.421 & 0.0523 &  9.823 & 41.3  &   \\
U3y04   & 2003 YT179 & 43.542 & 0.028  &  1.684 & 44.4  &   \\
U3h06   & 2003 HZ56 & 43.63  & 0.010  &  2.550 & 43.5  &   \\
U5c17PD & 1999 CN119  &  43.733 & 0.04043 &  0.999 & 44.5 & \\ 
U4n01 &   &  43.915 & 0.13500 &  0.271 & 43.7 & \\ 
U3y08   & 2003 YP179 & 44.03  & 0.079  &  0.947 & 41.3  &   \\
U4n02 &   &  44.056 & 0.06176 &  2.943 & 46.8 & \\ 
U3w16   & 2003 YN179 & 44.272 & 0.006  &  2.768 & 44.4  &   \\
U4j04PD & 2000 JF81  &  46.117 & 0.10218 &  1.742 & 44.9 & \\

\cutinhead{Scattering Disk}

%\loaddata{tables/ScatteringDiskNonCh}
  U7a01 &   &  42.621 & 0.16444 &  4.742 & 38.9 & I (5:3) \\

\enddata
\tablecomments
{$M:N$: object in the M:N resonance;
 I: indicates that the orbit classification is insecure
 (see \citet{2008ssbn.book...43G} for an explanation of the exact meaning).}

\end{deluxetable}

\NCKB\ of the tracked sample are in the classical belt,
split into \Ninn\ inner, \Nmain\ main, \Nout\ outer and \Ndetc\ detached belt
objects.
Orbital integration shows that
\NRes\ objects are in a mean-motion resonance with Neptune,
\NPlutino\ of which are plutinos.
The remaining sample consists of \NScat\ objects on scattering orbits.

The apparent motion of TNOs in our opposition discovery fields is
approximately $\theta (''$/hr) $\simeq$ (147 AU)/$r$,  where $r$ is the
heliocentric distance in AU.  With a typical seeing of 0.7~-~0.9~arc-second and a
timebase of 70~-~90~minutes between first and third frames, we were sensitive
to objects as distant as $r\simeq$125~AU, provided they are large enough to be
above our flux limit. The furthest object discovered in CFEPS lies at 58.3~AU
from the Sun (L3q03 = 2003~QX$_{113}$, a detached object with $a =
49.55$~AU).  The short exposure times used (70~-~90~seconds) allowed us to
detect objects as close as 15~AU without trailing. We elected to use a rate of
motion cut corresponding to objects further than 20~AU from Earth.

In the following sections we present a parameterization of 
the intrinsic classical Kuiper belt and scattering disk population implied by
our observations.
The differing detectability of these populations, in a flux-limited 
survey, implies that the intrinsic population ratios will be different
from the observed ones.  
We present the more complex
analysis of the resonant populations in a companion paper
\citep{2011AJ......1....1G}.

\section{The classical belt's orbital distribution}
\label{sec:model}

This section presents the results of our search for an empirical
parameterized orbit distribution for the various components of the 
so-called `classical' belt.
For each sub-component we start with a simple parameterization of the intrinsic 
orbit and absolute magnitude distributions.
We then use the CFEPS Survey Simulator\footnote{The survey simulator is
  available on-line, with all informations needed to use it, at {\it
    http://www.cfeps.net} as a stand-alone package, or at {\it
    http://CFEPSSim.obs-besancon.fr/}, as an on-line service.} to determine
which members of the intrinsic population would have been detected by the
survey.
The orbital-element distributions of the simulated detections are then compared
to our characterized sample.
This process is iterated with models of increasing complexity until arriving at
a model that provides a statistically-acceptable match; no cosmogonic
considerations are invoked.

Our model search process provided acceptable parameterizations of the main
classical belt, the inner classical belt and the outer+detached population.
Our goal is to discover the main features of the orbital distribution and
provide a population estimate for each orbital sub-component.  While our
success in finding acceptable models is not a proof of model uniqueness, we
were surprised, in many cases, by the restricted range of acceptable models.

To evaluate a model's quality, we extend the method defined in P1 to more
variables.  We compute the Anderson-Darling (AD) \citep{dvi898hb} statistic for
the distributions of the orbital elements $a$, $e$, $i$, $q$, and for $r$
(heliocentric distance at discovery) and $\mpg$ magnitude.  We use Kuiper's
modified Kolmogorov-Smirnov (KKS) statistic for the mean anomaly $M$.  We
follow the same procedure as used in P1 to determine the significance of the
computed statistics.  For each model parameterization we use the Survey
Simulator to draw a large `parent' population from the model.  We then draw
sub-samples with the same total number as in our L7 characterized sample.
Using this simulated `observed' population we compute the various statistics
that result from comparing to our large `parent' population.  This re-sampling
is repeated 5000 times providing a distribution of statistic values for the
given parameterization, ie. `boot-strapping' the statistic.  The probability of
statistic measured for the L7 sample is determined by comparing that statistic
value to the range of statistic values returned by the bootstrap process.  We
reject a model if the minimum statistical probability determined in this way is
returned by fewer than 5\% of the model bootstraps.

\subsection{The main classical belt}
\label{sec:mainbelt}

In P1 we presented a model that matched the orbital distribution of the main
classical belt objects detected in the L3 sample; due to the smaller number of
objects in the L3 sample, we restricted ourselves to fit only selected orbital
elements and considered a constrained range of the phase-space volume available
to main-classical belt objects.  In addition, P1 did not attempt to determine
the absolute magnitude distribution using our detections but instead utilized
values available in the literature.  Here we restrict our main classical belt
model to the 40~AU~$\le a \le$~47~AU range, to avoid the complex borders of the
3:2 and 2:1 MMR regions, which includes 88 characterized CFEPS TNOs.  
This sample size allowed us to remove external constraints on the magnitude 
distribution and explore a more complete model of the available phase-space.

\begin{figure}[h]
\begin{center}
\includegraphics[width=\columnwidth]{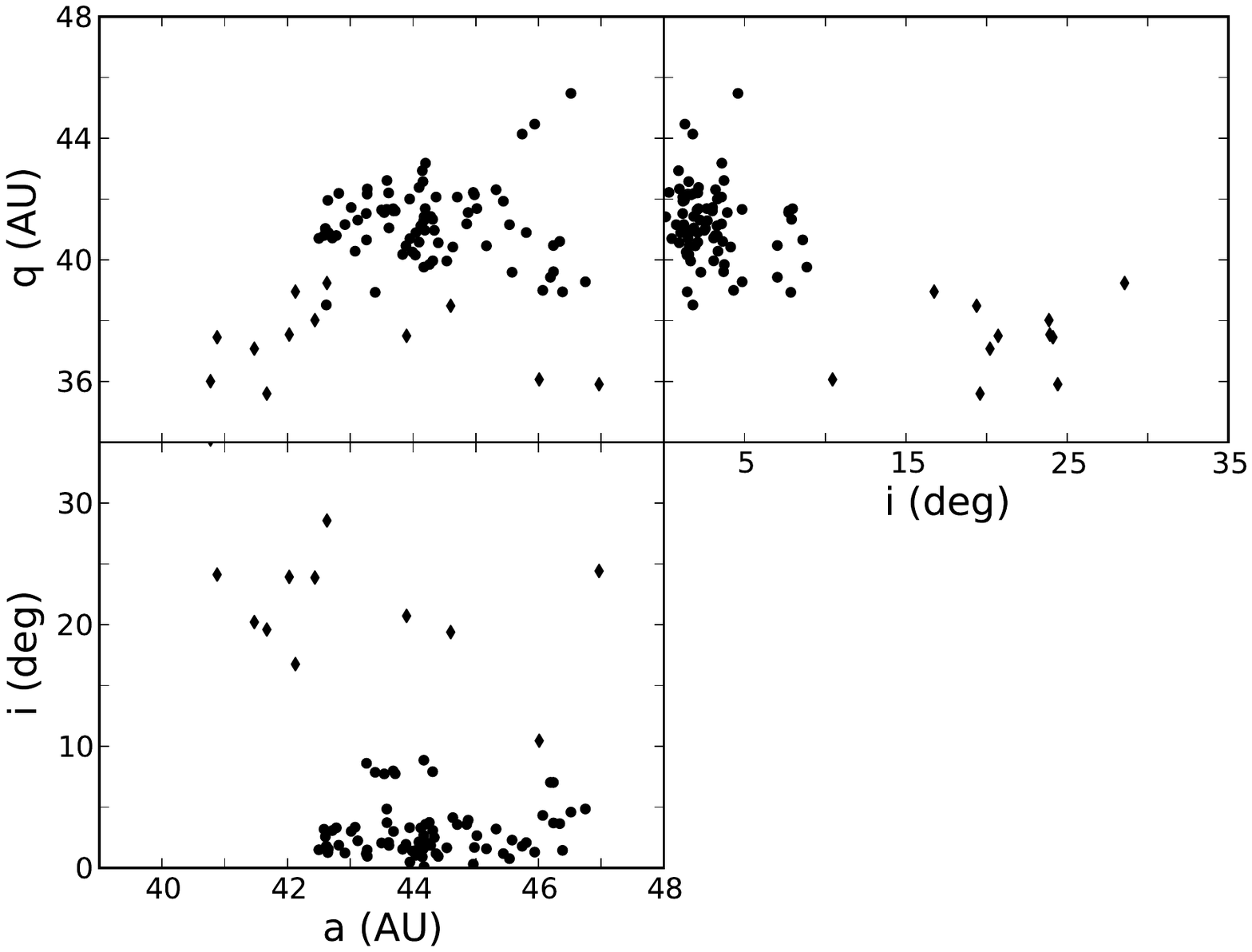}
\end{center}
\caption{Multiple 2D projections of ($a$, $q$, $i$) orbital elements of the
  CFEPS main classical belt objects. ($a$, $q$): upper left; ($a$, $i$): lower
  left; ($i$, $q$): upper right. Solid circles are for objects with $i <
  10^\circ$. Solid diamonds are for $i \ge 10^\circ$. 
  This cut is introduced to
  allow identification of large-$i$ TNOs in the $(a,q)$ plot,
  but has no relevance to our model.
}
\label{fig:CFEPS}
\end{figure}

\begin{figure}[h]
\begin{center}
\includegraphics[width=\columnwidth]{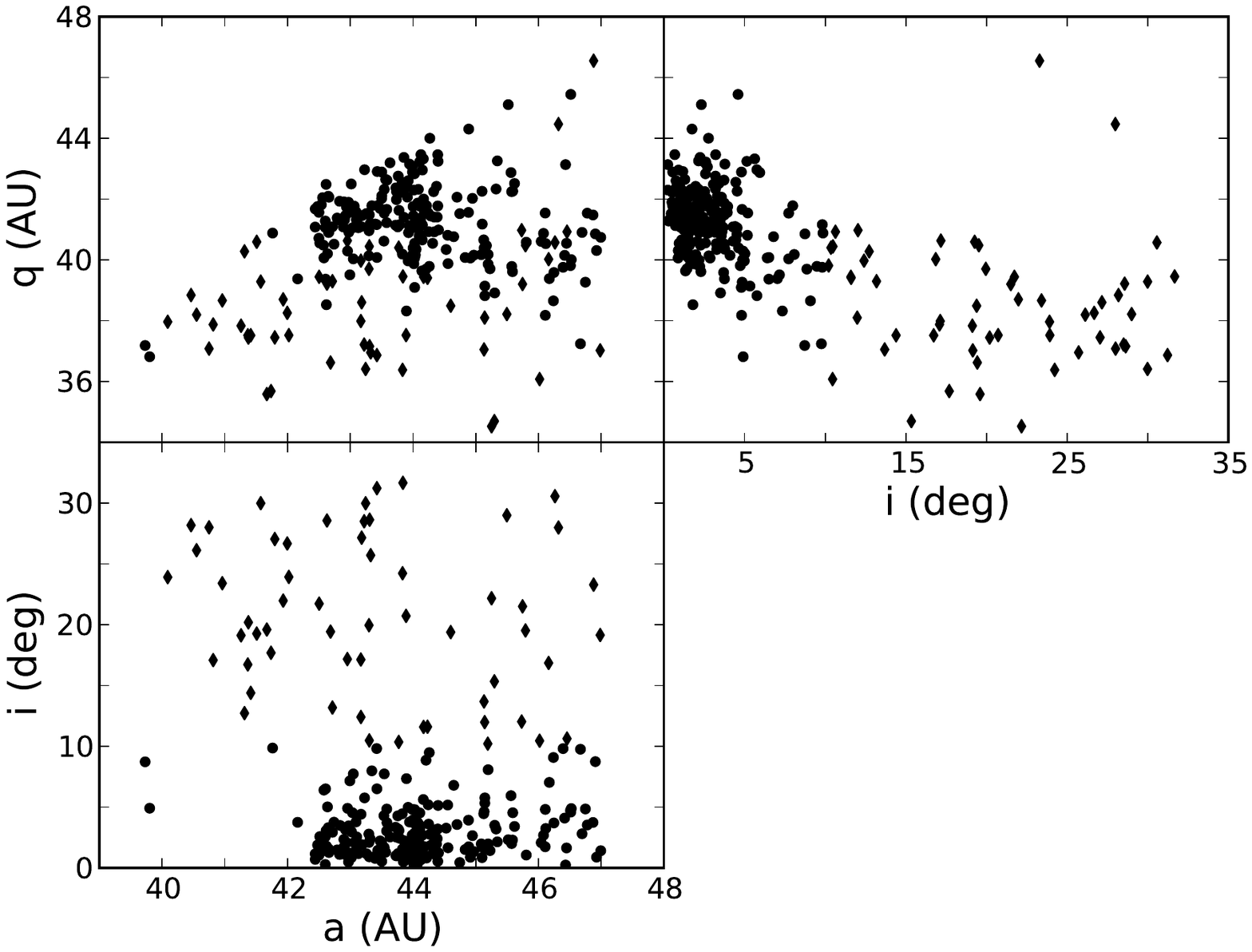}
\end{center}
\caption{Same as fig.~\ref{fig:CFEPS} but for the MPC main classical belt
  objects. L3 and Pre-survey objects are present on both plots, as
  well as any 'PD' object.
}
\label{fig:MPC}
\end{figure}

Figures \ref{fig:CFEPS} and \ref{fig:MPC} present ($a$, $i$), ($a$, $q$) and
($i$, $q$) projections of the main-belt TNO orbital elements for characterized
CFEPS detections and multi-opposition orbits in the MPC.  
These figures make it clear
that objects with $q<39$~AU are dominantly from the
high-inclination population, as was already apparent in the L3 model.  The
distribution of low-$i$ objects, which span a narrower range of semimajor axis
than their high-$i$ cousins, exhibit considerable phase space structure.  In an
effort to find a parameterization that yielded these interesting sub-structures
we investigated a substantial range of empirical representations.  We were,
however, unable to find a two-component 
model (like that in P1)
that sufficiently reproduced structure observed in the current sample.  
A more complex representation is required.

After much effort we arrived at our `L7 model' (based on CFEPS discoveries up
to mid-2007).  The L7 model is composed of three components
(Fig.~\ref{fig:L7mainModel}), the fine details of which are presented in
Appendix~A.  
These components are a population with a wide inclination
distribution (the {\it hot} population) superposed on top of a population with
narrow inclination component with two semi-major axis / eccentricity
distributions (the {\it stirred} and {\it kernel} populations).  
The hot
population is defined as a band in perihelion distance $q$ essentially confined
to the range 35 to 40~AU, with soft exponential decay outside this range.  
Using a `core' \citep{2005AJ....129.1117E} definition based only on 
inclination does not take into account the transition in the $e/i$ 
distribution beyond 
$a\simeq$44.4~AU clearly visible in both Figs.~\ref{fig:CFEPS} 
and \ref{fig:MPC}.  
With the qualifier that there will be mixing from the low-$i$ tail
from the hot component, we thus split the `cold' population of
the main classical belt into two sub-components.
The {\it stirred} population
have orbits drawn from a narrow-inclination distribution with semi-major axes
starting at $a$=42.5~AU and extending to $a \simeq 47$~AU, with a range of
eccentricities that increases as one goes to larger $a$.  
The stirred component does not contain
the sharp density change  at $a\simeq 44.5$~AU.  
There are more low-$i$ and moderate-$e$ TNOs per unit semimajor axis 
at $a \sim 44 - 44.5$~AU than at smaller and larger semi-major axis, 
indicating that a third component is required.  
To model this component we insert a dense
low-inclination concentration, which we call the {\it kernel}, near $a$=44~AU
to account for this intrinsic population.

The kernel may be the same as the clustering in the $a=$42--44 region seen as
far back as \citet{1995AJ....109.1867J} and \citet{1996AJ....112.1225J}.  This
also appears to be the same structure that \citet{2002ApJ...573L..65C} and
\citet{2003EM&P...92...49C} posited (with rightful skepticism) as a possible
collisional family.  Although we share the concern that normally the relative
speeds from a large parent-body breakup should be larger than this clump's
observed dispersion, we find that regardless of interpretation, there is
considerable observational support for a tightly-confined structure in orbital
element space near the location \citeauthor{2003EM&P...92...49C} pointed to.
Recent collisional modeling studies \citep[eg.][]{2010ApJ...714.1789L} raise
the possibility of grazing impacts forming low-speed families in the Kuiper
Belt, motivated by the Haumea family \citep{2007Natur.446..294B}. The large
number of $D \ge 170$~km (absolute magnitude\footnote{The g-band apparent
  magnitude of a TNO at heliocentric and geocentric distance of
  1~AU if viewed at 0$^\circ$ phase angle} $H_g \le 8$) objects in the kernel
implies that the parent body would have been a dwarf planet at least as large
as Pluto, an unlikely possibility.  The kernel thus appears to be the
longest-recognized dynamical sub-structure in the classical Kuiper Belt,
a structure which requires confinement in all of $a$, $e$, and $i$.

There may be other possible representations of the orbital distribution that
are consistent with the CFEPS detections, with different boundaries or
divisions of the phase space.  We have found, however, the generic necessity of
a 3-component model can not be avoided.
The main characteristics of our model must be similar to reality,
because a considerable amount of tuning was needed to achieve an acceptable
model.
From this 3-component model, we can then provide robust measurements of the
sizes of the subpopulations in the Kuiper belt and generate a synthetic
`de-biased' model of the orbital distribution of the main belt which can be
used for various modeling purposes, such as collisional dust
production \citep{2010AAS...21534606S}.

\begin{figure}[h]
\begin{center}
\includegraphics[width=\columnwidth]{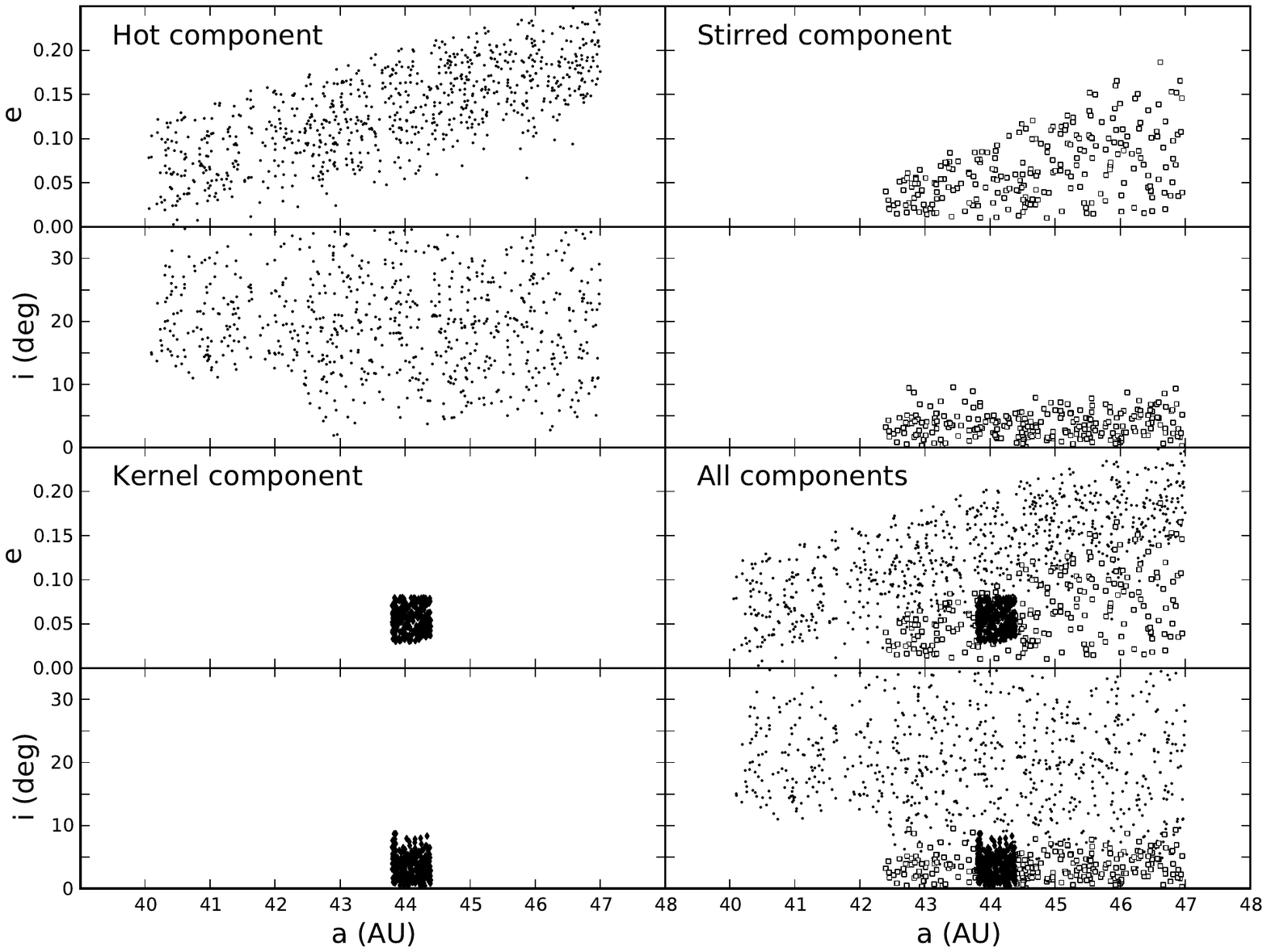}
\end{center}
\caption{The three components of the CFEPS-L7 synthetic model for the main
  classical belt. The hole at low-$i$, low-$a$ in the hot component is
  introduced to represent the destabilizing action of the $\nu_8$ secular
  resonance.
}
\label{fig:L7mainModel}
\end{figure}

\subsubsection{The luminosity function}
\label{sec:lumfun}

The absolute magnitude $H_g$ distribution can be represented by an exponential
function $$N({\rm H}) \propto 10^{\alpha {\rm H}} \label{eq:hdistrib}$$ with
`slope' $\alpha$.  $H_g$ is converted into apparent magnitude $\mpg$ by $\mpg =
H_g + 2.5 \log{(r^2 \Delta^2 \Phi(\mu))}$, where $\Delta$ is the geocentric and
$r$ the heliocentric distance, $\mu$ the phase angle (Sun-TNO-observer) and
$\Phi(\mu)$ the phase function defined by \citet{1989aste.conf..524B}.  We find
that two different values of $\alpha$, one for the hot and one for the cold
distributions, are required by our observations.  Allowing the stirred and
kernel components to have differing values of $\alpha$ did not provide an
improved match to the observations and is not required.

We have run a series of model cases using the orbital element distributions
described previously while varying the luminosity functions slopes for the hot
component, $\alpha_h$, and for the cold (kernel + stirred) components
$\alpha_c$.  For each case, we varied the other orbit model parameters to find
the best possible match between the cumulative distribution functions of the
Survey Simulator observed Kuiper belt and the L7 sample for each of the
selected values of $\alpha_h$ and $\alpha_c$.  In this way we determined the
range of allowed power-law slopes for the limited range of TNO sizes, $7
\lesssim H_g \lesssim 8 $, probed by our observations.

\begin{figure}[h]
\begin{center}
\includegraphics[width=8cm]{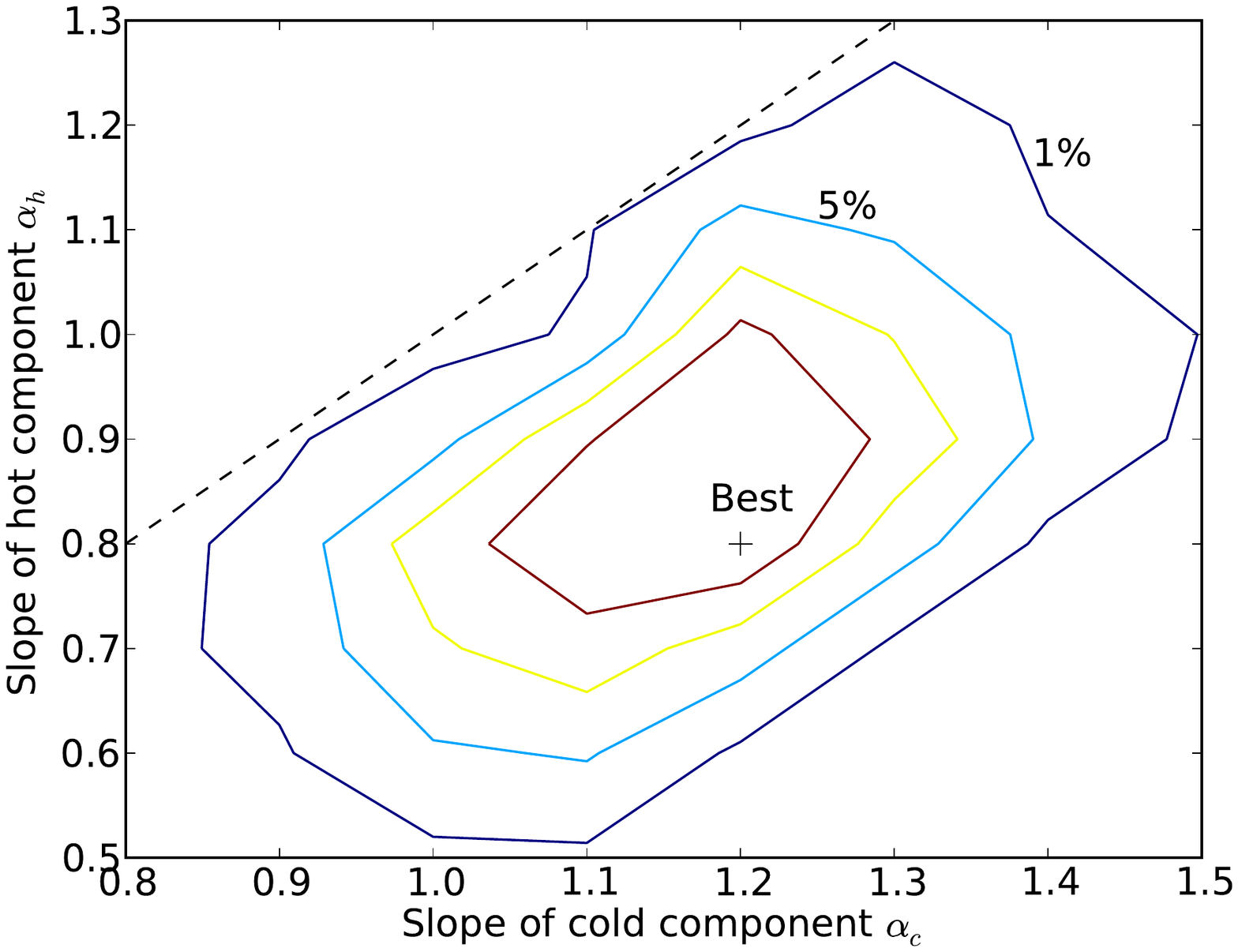}
\end{center}
\caption{Contour plots of the `minimum probability' statistic for a range of
  main classical belt models. Each model has a different slope of the $H$
  distribution for both the hot component ($\alpha_h$) and the other components
  ($\alpha_c$). Contour levels for 1\% and 5\% probabilities are
  shown. Acceptable models are interior to the 5\% level curve.  The dashed
  line indicates the locus with identical slopes for all components.  Th e plus
  sign indicates the adopted model (which gave the best match).
}
\label{fig:slopes}
\end{figure}

Our best fit values are $\alpha_c = 1.2^{+0.2}_{-0.3}$ and $\alpha_h =
0.8^{+0.3}_{-0.2}$, with Fig.~\ref{fig:slopes} presenting the joint 95\%
confidence region for these slopes.  A single value of $\alpha$ for all
sub-components in our model is rejected at $>$99\% confidence.  The $\alpha_c$
determined here is in good agreement with the range derived by
\citet{2004AJ....128.1364B} for the low-inclination objects and somewhat
steeper than that reported in \citet{2005AJ....129.1117E} while our value for
$\alpha_h$ overlaps the ranges proposed by both \citet{2004AJ....128.1364B} and
\citet{2005AJ....129.1117E} for what they call the excited population.
\citet{2010Icar..210..944F} also found markedly different values for the 
slope of the cold component, 0.59--1.05, and the hot component, 0.14--0.56. 
While those
slopes are consistent with \citet{2005AJ....129.1117E} they are shallower than
\citet{2004AJ....128.1364B} and our own estimates. The
\citet{2010Icar..210..944F} results, however, probed smaller-size objects than
our observations and the difference in slopes may be reflective of a change in
size distribution around $H\sim8.5$ where the CFEPS detections dwindle.  Thus,
in the limited size ranges probed by these surveys, there appears to be
reasonable agreement on the slope of luminosity function for these components
of the Kuiper belt with the hot and cold components exhibiting slopes that are
significantly different.

The value of size distribution slopes reported here range from 0.8--1.2 and are
considerably larger than the best-fit slopes discussed in many previous
analyses that attempted to determine a global luminosity function for the
Kuiper belt.  For example, \citet{2008ssbn.book...71P} reviewed estimates of
$\alpha$ ranging from 0.5--0.8 for surveys that cover the range $H_g
\simeq$5--10.  \citet{2009AJ....137...72F} and \citet{2009ApJ...696...91F}
demonstrated that a slope of $\sim \alpha=0.75$ is a decent representation of
the `average' belt down to magnitude $\sim m_r$=25, but that there is a gradual
flattening of the apparent luminosity-function slope at fainter magnitudes,
continuing to a slope which may become extremely flat somewhere beyond $H>10$
according to the \citet{2004AJ....128.1364B} analysis of a deep HST search.
The quest for a single `master' luminosity function, however, is misguided:
\begin{enumerate}
\item Because there are different slopes for the hot and cold main-belt
   components, the slope should be $\alpha\simeq$0.8 at large sizes
   (where the hot component dominates) and become steeper (if looking in
   the ecliptic where the cold population is visible) when the depth of
   the survey results begins to probe the size range at which the
   cold-population surface density becomes comparable to the hot population.
\item The on-sky density of the (essentially non-resonant) cold population is
   essentially dependent only on the ecliptic latitude. The hot population's
   sky density varies with both latitude and longitude due to the fact that
   the resonant populations are hot.  Thus, the $H$ magnitude at which the
   steeper cold component power-law takes over will also depend on the latitude
   and longitude of the survey.
\end{enumerate}

Interestingly, extrapolating from the $\sim$4000 objects in the cold belt with
$H_g \le $8  (see sect.~\ref{sec:popest}) to larger objects, one finds that
there should be only $\sim$1 TNO with $H_g < $5. 
This is consistent with the current census of large objects in the cold belt, 
which should be close to complete \citep{2003EM&P...92...99T}.
Similarly, one would expect to have only $\sim$1 TNO with $H_g < $3.5 in the
non-resonant hot population, which again corresponds to our knowledge of the
Kuiper belt \citep{2008ssbn.book..335B}.
Currently, the MPC report 6 objects with absolute magnitude $ < $3.5 in the
classical belt region as defined for our population estimate. 5 of them are
clearly part of the hot population, with inclinations between 20 and 30
degrees, the last one being Quaoar with an intermediate inclination of 8
degrees.

The realization that the hot component has a low-$i$ tail means that caution
must be exercised because one simply cannot isolate the `cold' cosmogonic
population with the commonly-used $i<5^\circ$ cut.  For example, in the
ecliptic at bright (say roughly $m_r\sim 22$) magnitudes, the low-$i$ tail of
the hot component can be numerically comparable to the sky density of `cold'
objects.  Thus, it is not possible to isolate the cold component at bright
magnitudes based simply on orbital inclination.

\subsubsection{Acceptable range for main parameters.}
\label{sec:bestfit}

In this section we fix the slopes just determined, i.e. $\alpha_h = 0.8$ and
$\alpha_c = 1.2$ and examine the range of model parameters allowed by the L7
detections.  Due to the large number of orbital parameters to adjust and the
time required by each survey simulation (10--50 minutes on the fastest
available computers), we did not run an automated minimum-finding algorithm,
but rather did a manual search on a multidimensional parameter grid.

Acceptable values (rejectable at less than 95\% confidence) for the inclination
width (see Appendix~A) of the hot component $\sigma_h$ range from
14$^\circ$--29$^\circ$.  A hot-component width $\sigma_h = 16^\circ$ is
acceptable not just for the main-belt population but also reproduces the
observed inner and outer classical populations (see Sections \ref{sec:inner}
and \ref{sec:outer}) and thus we adopt this value as the width of hot
component.

The acceptable range for $\sigma_c$ is 2.3$^\circ$-3.5$^\circ$, with a peak of
the probability near 2.6$^\circ$, which we adopt.  
\citet{2001AJ....121.2804B}
analysed the MPC database at the time and concluded the existence of the cold
component to the inclination distribution; with $\sigma_c = 2.2^{+0.2}_{-0.6}$
degrees (one-sigma uncertainties), consistent with our results.
\citet{2005AJ....129.1117E} in their initial analysis of the Deep Ecliptic
Survey estimated a $1.94 \pm 0.19$-degree width for the cold component.
\citet{2010AJ....140..350G}, however,
recently re-analysed the detections from the Deep Ecliptic Survey, and found a
$2.0^{+0.6}_{-0.5}$-degree width (one-sigma uncertainties) for the cold
component.  Thus, the DES is also in reasonable agreement with our results,
given the uncertainties.
\citet{2004AJ....127.2418B} found a much narrower width of $1.3^\circ$ (no
uncertainty given) for the cold component, with respect 
to a locally-determined Laplace plane for each semimajor axis.
We have not repeated a similar analysis.

The L7 distribution contains an excess of
intermediate-inclination objects ($i$ in range 6$^\circ$--10$^\circ$) when
compared to models with $\sigma_h \ge 16^\circ$ and $\sigma_c = 2.2^\circ$.
This `bump' in the cumulative inclination distribution can also be seen in the DES
sample, \citet[][Fig.~13]{2002AJ....123.2083M} and
\citet[][Fig.~17]{2005AJ....129.1117E}, between inclinations of
8$^\circ$ and 10$^\circ$.
The Survey Simulator approach accounts for the distributions of all orbital
elements simultaneously and thus the L7 model makes 
the inclination bump part of the cold component because the objects in this
inclination range have $e$ and $a$ distributions that
make them part of the cold component, hence increasing its width.
Although it was possible to keep a cold width of
2.2$^\circ$ or lower by introducing a third inclination component the
observations do not currently demand this increase in complexity.

The observed cumulative inclination distribution has two steep increases
corresponding to the cold and hot component, both of which are steeper than for
our model.  This indicates that the actual differential distribution of each
component is probably more confined than $\sin(i)$ times a Gaussian centered on
zero.  It is remarkable that the hot component of the main classical belt
extends up to 35$^\circ$ and stops abruptly.  This limit is seen not only in
the CFEPS, but also in the MPC databases (see Figures~\ref{fig:CFEPS} and
\ref{fig:MPC}).  We experimented with $\sin^2(i)$ times a Gaussian centered on
zero, but this did not result in a significant improvement to our fit.  Note
that \citet{2005AJ....129.1117E} find that $\sin(i)$ times a Gaussian plus
Lorenzian give their best fit to the classical belt inclination distribution.
More recently, \citet{2010AJ....140..350G} find that $\sin(i)$ times a Gaussian
of width $\sim 7^\circ$ and centered around $\sim 20^\circ$ best fits what
they call the `Scattered Object' inclination distribution.  We did not test this
functional form as this introduces an extra parameter, which is not demanded by
the current sample.  The \citet{2001AJ....121.2804B} functional form may not be
an exact representation of every sub-component's inclination distributions; we
can, however, obtain an acceptable match to the CFEPS survey with this
functional form.

The fraction of each component (hot versus cold inclination components) varies
with the $H_g$-magnitude limit, due to their differing values of $\alpha$.  We
report here the acceptable range for the fractions of each sub-population at
the $H_g \le $8.0 limit.  We find that the fraction of the hot component,
$f_h$, cannot exceed 62\% and is at least 33\%, with a best match to the
observations at $f_h \simeq 0.51$.  This hot-component fraction and widths are
close to the nominal L3 model from P1.  We find that the fraction of the kernel
$f_k$ has to be larger than 0.05, but less than 0.30 at 95\% confidence and
adopt $f_k = 0.11$.  The fraction in the stirred is then $f_s$=0.38, when
considering $H_g<8.0$.

\begin{figure*}[h]
\begin{center}
\includegraphics[width=18cm]{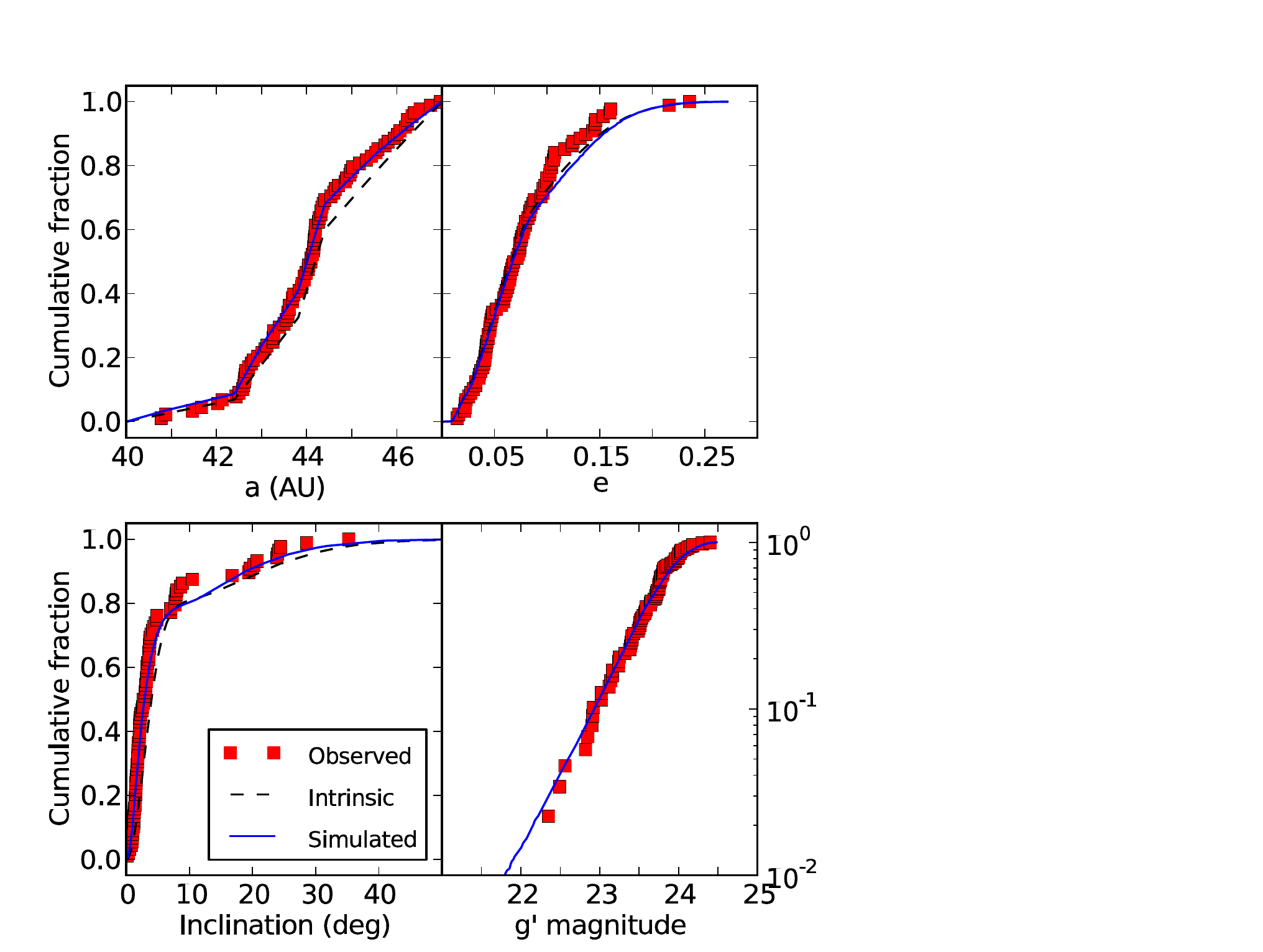}
\end{center}
\caption{CFEPS+Pre objects (red solid squares) compared to our main-belt
  model's distribution in $a$ (upper left), $e$ (upper right), $i$ (lower left)
  and $\mpg$ magnitude (lower right) distributions when the intrinsic (dashed
  line) are observed through the CFEPS Survey Simulator (resulting thin solid
  lines). The model used here is the one described in
  Section~\ref{sec:mainbelt}, with values of the parameters
  corresponding to our nominal case (see Section~\ref{sec:bestfit}).
}
\label{fig:cdfs1}
\end{figure*}

Figure~\ref{fig:cdfs1} presents the comparison of our nominal model with $a$,
$e$, $i$ and $\mpg$ apparent-magnitude distributions.  When biased by the CFEPS
survey simulator, the L7-model reproduces the detections extremely well.

Our hot/cold population fractions differ from those reported in some 
other works, but details are important in the comparison.
\citet{2001AJ....121.2804B} report a hot fraction of 81\%.  This fraction
listed must be treated with the caution engendered by the realization that the
MPC sample has a non-uniform $H$-magnitude limit, making interpretation of a
fractional population (given the different luminosity functions) difficult. The
\citet{2010AJ....140..350G} estimate is even more difficult to compare, 
because the classification scheme used explicitly separates out many of the 
highest-inclination main-belt TNOs into portions of the 
`scattered' population (even though many of these TNOs are very decoupled 
from Neptune) and thus the relatively small `hot'
width of $8^{+3}_{-2}$ has been forced down;
a direct comparison of the relative populations is thus not possible.
\citet{2001AJ....122..457T} has a $H$-magnitude limit that is more uniform than
the MPC sample but they mix together the various orbital classes when
reporting the relative fraction of hot and cold component objects.

\subsubsection{Population Estimates}
\label{sec:popest}

The procedure in Section~4.3 of P1 was used to derive a population 
estimate for the main classical belt. 
Unlike much of the literature, which gives population estimates for
objects larger than an estimated diameter, 
CFEPS gives population estimates for absolute magnitude smaller than a 
given value of H$_g$, and thus an unknown albedo is not introduced
into the estimate\footnote{More subtly, surveys at different latitudes
and longitudes probe different average distances as they look into the
trans-neptunian region due to the different distance distibutions of
resonant and non-resonanat populations; thus a given apparent magnitude
depth actually probes at different average size limit.  Stating a 
population limit to a stated $H$-magnitude limit is thus more 
meaningful.}.
These estimates and their uncertainties are given assuming our orbital
model. They would change if we were to change our parameterization. 
In particular, increasing the width of the inclination
distributions `hides' more of the population far from the ecliptic. 
Alternately, decreasing the cold component's width to $1.3^\circ$ requires 
changing the hot/cold fraction and results in a decrease of the 
total main-belt population by 20\%.

In principle, the very
deepest blocks in our survey are sensitive to a limit of $H_g\simeq9.5$
for a perihelion detection on the most eccentric orbits in our main-belt 
model.
The Survey Simulator shows, however, that based on our model orbit 
distribution, the vast majority of our detections should have 
H$_g \le 8.0$, consistent with our largest-$H$ classical-belter 
detection, H$_g = 8.1$.  
Thus our population estimate is given to the limit to which the survey
has reasonable sensitivity:
\begin{displaymath}
N_{\rm classical} ({\rm H}_g \le 8.0) = ( 8000^{+1800}_{-1600} )
\end{displaymath}
\noindent
where the uncertainties reflect a 95\% confidence limit assuming the underlying
orbital model and its parameter values are correct. Our measured value for
$\alpha$ essentially is only for the range $H_g = 7$--8 which dominate our
detections.

\begin{figure}[h]
\begin{center}
\includegraphics[width=\columnwidth]{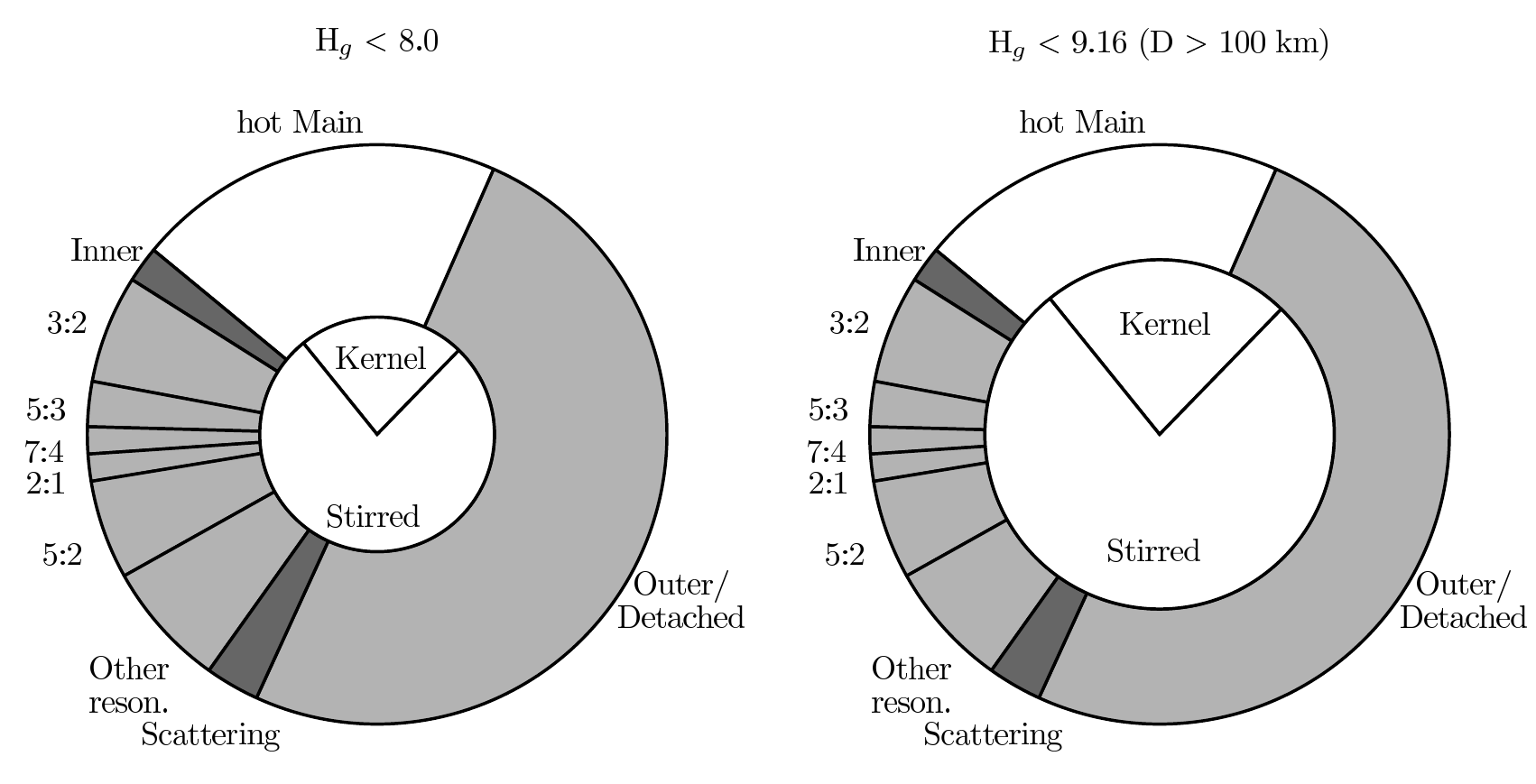}
\end{center}
\caption{A representation of the fractional populations of the various
  dynamical classes measured in the L7 model. The surface area of each
  population shown is proportional to the relative population for objects with
  $H_g \le 8$ (left) and $H_g \le 9.16$ (right), corresponding to $D \ge
  100$~km, assuming an albedo $p=0.05$.  The wedge label ``Other reson.''
  refers to resonant populations measured other than those individually labelled
  \citep[4:3, 7:3, 5:4, 3:1, 5:1; see][]{2011AJ......1....1G} .  The outer
  annulus is comprised entirely of ``hot'' population objects while the
  ``cold'' populations, of which there is only the Kernel and Stirred
  components, are represented by the inner circle. The white area corresponds
  to the main belt.
}
\label{fig:schematic}
\end{figure}

The formula
$$
N({\rm H}_g \le {\rm H}_1) = 10^{\alpha \times \Delta H} N({\rm H}_g \le {\rm H}_0) \; ,
$$
\noindent
where $\Delta H = {\rm H}_1 - {\rm H}_0$, allows one to scale population
estimates of P1 to H$_g = 8.0$, and also compare with other populations like
the inner belt or the plutinos (which can come closer to Earth than the main
classical belt).  Here care must be taken to distinguish between the hot
components and the others because they have different H-magnitude slopes, hence
the extrapolation factor to any particular $H$-limit is different for each
sub-component.  Figure~\ref{fig:schematic} shows a schematic representation of
the fractional population sizes of all the dynamical classes measured in the L7
model.  
This figure demonstrates that one must be careful when comparing the
relative sizes of various sub-populations whose size-distributions are
different because the relative populations will vary with the 
H-magnitude limit being considered.

Due to the lack of phase relations with Neptune and good statistics due
to large numbers of main-belt detections in ecliptic surveys, the
main classical-belt population estimates should be the most certain in the
literature of all the populatoin estimates. 
Comparing the L7 main-belt estimate with the literature yields 
satisfactory agreement (details are given in Appendix~C).
Table~\ref{tab:pops} provides our current population estimates, after
accounting for the size distribution scalings and using the 
same assumptions as in P1, i.e. an albedo of $p_g$=0.05, hence
H$_g$(D$_p$=100~km)~=~9.16.
\citet{2005AJ....130.2392H} give an essentially-identical estimate of 
130,000 TNOs with $40.1<a<47.2$~AU and $D>100$~km (with no error 
estimate), which is certainly within our 95\% confidence region even
with the small differences in albedo and phase-space boundaries 
used.
If we use constant $\alpha$ values for the two components 
and extrapolate to $D > $100~km, we find the same population estimate
as \citet{2005AJ....130.2392H} and are a factor of a few higher than
\citet{2001AJ....122..457T} (see Appendix~C). 

Our current estimates agree with P1 when scaled to the $H_g<8$ limit
where CFEPS is sensitive.
Thus, it appears that the main belt's population for $H_g<8$ 
is secure, where we have provided the first detailed breakdown of the
hot and cold component's individual populations and detailed
sub-structure.

\begin{deluxetable}{lcc}
\tabletypesize{\footnotesize}
\tablecolumns{3}
\tablewidth{0pc}
\tablecaption{Model dependent population estimates.%\\
\label{tab:pops}}
\tablehead{
\colhead{\ } & \colhead{$N({\rm H}_g \le 8)$} &
\colhead{$N({\rm D} \ge 100{\rm km})$}% \\
}
\startdata
\cutinhead{Inner Classical Belt}
All             & $   400^{+  400}_{-  200}$ & $  3,000^{+ 3,500}_{- 2,000}$ \\
\cutinhead{Main Classical Belt}
Hot             & $ 4,100^{+  900}_{-  800}$ & $ 35,000^{+ 8,000}_{- 7,000}$ \\
Stirred         & $ 3,000^{+  700}_{-  600}$ & $ 75,000^{+17,000}_{-15,000}$ \\
Kernel          & $   900^{+  200}_{-  200}$ & $ 20,000^{+ 5,000}_{- 5,000}$ \\
All             & $ 8,000^{+2,000}_{-2,000}$ & $130,000^{+30,000}_{-27,000}$ \\
\cutinhead{Outer/Detached Classical Belt}
All ($a>$48)    & $10,000^{+7,000}_{-5,000}$ & $ 80,000^{+60,000}_{-40,000}$ \\
\enddata
\tablecomments
{Our model estimates are given for each sub-population within 
the Kuiper belt. The uncertainties reflect 95\% confidence
intervals for the model-dependent population estimate.
Values for $N($D$>$100~km$)$ are derived assuming an albedo of $p_g$=0.05,
hence $H_g = 9.16$.
Remember that the relative importance of each population
will vary with the upper $H_g$ limit.
}
\end{deluxetable}

\subsubsection{Discussion}

Some characteristics of the main classical belt that require explanation are
the bimodal nature of the inclination distribution, the relative importance of
the so-called hot component, and the marked sub-structures in $(a,e)$ space for
the low-inclination objects.

\citet{1998AJ....115.2125J}, \citet{2001AJ....122..457T},
\citet{2001ApJ...549L.241A}, \citet{2001ApJ...554L..95T} and
\citet{2008ssbn.book...59K}, reported the existence of an edge of the Kuiper
Belt at 47--50~AU.
Because the samples on which they based their estimate were
heavily biased towards low-inclination objects, they were really detecting an
edge of the cold component of the classical belt.  In addition, Figure 14 of
\citet{2001AJ....122..457T}, Figures 2 and 3 of \citet{2001ApJ...554L..95T},
and Figure 3 of \cite{2008ssbn.book...59K} all show a marked peak at around
44~AU followed by a very fast decrease in the number of objects past
44.5--45~AU, with perhaps a low density tail past 50~AU. The above papers vary
in how sharp they consider the ``cut off'' to be.
In hindsight it is clear that what they were reporting as an edge is in fact
due the presence of the low-inclination kernel and stirred components, which
dominate the low-latitude detections and fall off quickly beyond 45~AU.  As
\citet{2008ssbn.book...59K} 
point-out, the peaked nature of the
distribution is absent in the 'hot' component and entirely absent from
the `scattering disk' population.  The stirred component's density is a
rapidly-decreasing function of semimajor axis that 
becomes very small by the time the 2:1 resonance is reached.
This hints at a possible connection between the kernel, the
stirred component and the migration of the 2:1 to its current location;
this outer edge appears only in the low-inclination component.  
We will show below that a scenario with the hot
component continuous across the 2:1 resonance is in agreement with the data.

The L7 sample contains a cluster of 6 objects with large $e$ and $i$ just
interior to the 2:1 MMR.  Amongst these, only the one with $a < 47$~AU
(L4k17, $a = 46.967$) was included in our analysis of models of the
classical belt, the other five being in the region where the exact limit of the
resonance is not easy to analytically define. This cluster could very well be a
group of objects ``dropped out'' when the 2:1 MMR shrank at the end of
Neptune's evolution (Sec.~\ref{sec:models}).

\citet{2006Icar..183..168G} reported a difference between the B-R color of the
`Core' and the `Halo', the former being redder than the latter, from
photometric measurement they later acquired on the DES sample. 
Our orbital survey was also not designed to yield precision photometry, and 
the $\mpg-\mpi$ and $\mpg-\mpr$ colors that we
can obtain from Table~\ref{tab:phot} are too uncertain to address this point.

In Sec.~\ref{sec:concl} we discuss some consmogonic implications of these
features, review how well the current models reproduce them, and propose future
directions.

\subsection{The inner classical belt}
\label{sec:inner}

The `inner' classical belt is the non-resonant and non-scattering population
between Neptune and the 3:2 resonance. Paper P1 contained only two such TNOs,
preventing us from deducing a detailed description of this region of the
Kuiper belt.
There are six inner classical belt objects in the L7 sample (see
Table~\ref{tab:chclass}), providing the opportunity to start constraining an 
orbital distribution.  The phase space is cut by the $\nu_8$ secular resonance
which eliminates almost all inner-belt TNOs with $7^\circ < i < 20^\circ$
making the intrinsic inclination distribution difficult to interpret.
If one uses 
a definition of `cold' belt as those objects with $i<5^\circ$ 
\citep[{\it eg.,}][]{2007Icar..189..213L},
one concludes that a large fraction of
the inner belt is cold.  Such an analysis, however,
neglects the bias towards detecting the lowest-$i$ TNOs from the hot population
in ecliptic surveys and the removal of moderate inclination objects via the
$\nu_8$.  Determining the intrinsic orbital distribution of the inner belt is
precisely the sort of problem in which a simulator approach provides a
clearer understanding.

\subsubsection{Parametric Model}

For the inner-belt population we utilized the same form of semi-major axis and
perihelion distance distributions as for the hot component of the main
classical belt (see Appendix~A), changing the range of semi-major axis to be
$37<a<39$~AU and fixing the size distributions for the hot and cold components
to be the same as those found for the main belt populations.  We then attempted
to find a model that included both a hot and a cold component using the same
inclination widths and fractions as for the main belt, these models were
rejected at $>$95\% confidence.  
Using the same $(a,q)$ model but with a
single-component inclination distribution width of $\sigma_h=16^\circ$, 
like the main belt's hot component (cutting away 
$7^\circ < i < 20^\circ$ orbits as they were proposed)
provides a perfectly-acceptable match to the L7 inner-belt 
detections.  
In fact, inclination widths of $5^\circ < \sigma_h < 20.0^\circ$ were 
found to be acceptable.
Even restricting one's attention only to the inner-belt TNOs with 
$i < 7^\circ$ (inclinations below
$\nu_8$ instability region) still requires an inclination distribution wider
than the cold component of the main belt, indicating that the evidence against
an inner-belt cold component comes from not just the largest-$i$ detections.

\subsubsection{Population Estimates}

Using a single component model with $\sigma_h = 16^\circ$ and $\alpha=0.8$ we
determine $N_{\rm inner} ({\rm H}_g \le 8.0) = 400^{+400}_{-200}$
(Table~\ref{tab:pops}).  This estimate is in good agreement with the
L3-sample's estimate of $290^{+690}_{-250}$.  As before, the uncertainties
reflect 95\% confidence limits given the intrinsic model distribution and does
not reflect our uncertainty in the model.  
These random uncertainties are a factor of two,
due to the small number of inner belt detection.

\subsubsection{Discussion}

\citet{2010AJ....140...29R} compared photometric colors of inner-belt TNOs to
other categories and found a good match between the inner belt and the
high-inclination objects from the main belt, while a marked difference from the
low inclination objects from the main belt, supporting the ``hot-only''
hypothesis.  To attempt to duplicate the conclusion, we compared our
photometric data for each population.  Unfortunately, but also unsurprisingly,
the quality of our photometric data is insufficient for such a comparison.  The
median uncertainty on our $\mpg-\mpi$ and $\mpg-\mpr$ colors is $\sim 0.25$,
which is about five times more than for the \citet{2010AJ....140...29R} data.
We are thus unable to provide additional verification from our 
current photometric colors.

\begin{figure}[h]
\begin{center}
\includegraphics[width=\columnwidth]{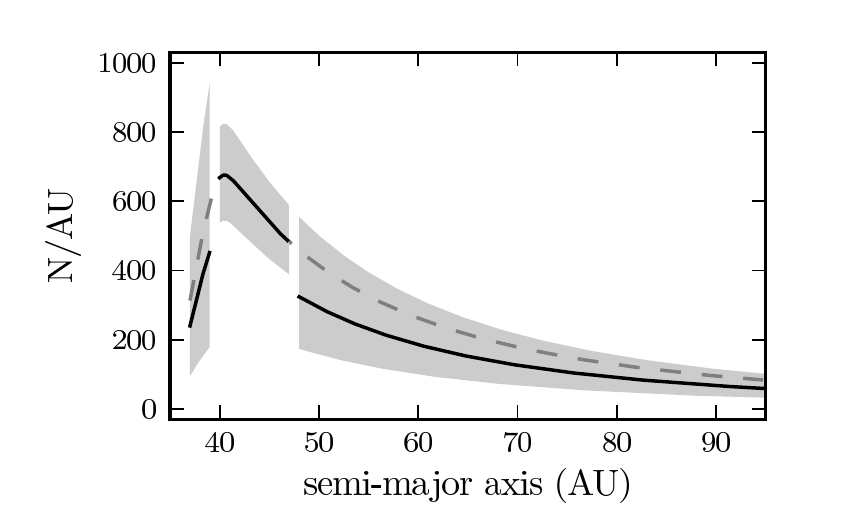}
\end{center}
\caption{The linear number density (/AU) for three Kuiper Belt components:
  the inner belt ($a < 39$~AU), the hot main belt ($40 < a < 47$) and the
  outer plus detached belts ($a > 48$~AU).  Each region's total population
  is scaled to the number with $H_g \le 8$, as determined by our model 
  population estimates.  The inner belt's population has been scaled up by 
  a factor of 1.85 to account for 
  the $\nu_8$ resonance (see footnote \ref{fn:innernu8}). 
  The solid lines represent
  the model population determined {\it independently} for each zone while 
  the grey
  dashed line indicates the smooth extension of the hot main belt model to the
  semi-major axis range occupied by the inner belt and the outer+detached
  populations,  
  where the inner-belt decay at lower $a$ occurs because of the 
  rapidly-shrinking stable $(a,q)$ phase-space volume available.
   A continuous primordial $a^{-2.5}$ hot population could, within
   uncertaties, account for all three populations.
  This suggests that these three Kuiper Belt components 
  are a single dynamical population.
}
\label{fig:innerHotOuter}
\end{figure}

The successful use of the same orbital distribution for the inner belt and the
main-belt's hot component suggests that the entire inner belt may be the
low-$a$ tail of the hot main belt.  This would be a cosmogonically appealing
unification of the sub-populations of the Kuiper Belt.  If true, then (at least
to order of magnitude) the TNO linear number density at the boundary 
(we chose 40~AU) 
extracted from each model should be comparable.
Denoting $P({\rm H}_g < 8.0)$ as the number of objects per AU with ${\rm H}_g <
8.0$, we find $P_{\rm inner} ({\rm H}_g < 8.0, 40\;{\rm AU}) =
270^{+180}_{-100}\, {\rm AU}^{-1}$.  For the hot main belt, $P_{\rm main} ({\rm
  H}_g < 8.0, 40\;{\rm AU}) = 670^{+160}_{-140}\, {\rm AU}^{-1}$.  At this
interface, the hot main-belt number density is $\sim3$ times that of the
extrapolated inner belt.  Given the very uncertain nature of these estimates
and the fact that they are anywhere close leads us to postulate that the
inner-belt and hot-main TNOs were emplaced by a single cosmogonic process.  In
this hypothesis, the reduced inner-belt density would be due to the smaller
volume of stable phase space in the inner belt region (because there is a
smaller available stable range of $e$) as well as the significant range of
inclinations from 7$^\circ$ to 20$^\circ$
destablized by the $\nu_8$ .
Scaling the inner-belt population density, to account for this reduced
inclination range, results in $P_{\rm inner}({\rm H}_g \le 8.0, 40\; {\rm AU}) =
500^{+300}_{-200}\, {\rm AU}^{-1}$, consistent with the value from the main
belt estimate at the $2\sigma$ level.  Figure~\ref{fig:innerHotOuter} presents
the linear number density versus $a$ for the scaled inner belt\footnote{The 
7-20$^\circ$ portion of a $\sin i \exp(-0.5i^2/(16^{{\circ}2})$ 
inclination distribution accounts for 46\% of the $\sin i$-weighted
phase space; to correct a population in the remaining phase space
back to the original needs to be multiplied by 1/(1-0.46) = 1.85.
\label{fn:innernu8}
}
compared to those of hot main belt and outer+detached
populations.

In this picture, the lack of a cold inner-belt component is significant.
Assuming that the cold component originally existed in this region, the
plausible mechanism for the cold component's destruction is the $\nu_8$
resonance sweeping out through the inner belt at some time, eliminating all
low-$i$ TNOs.
The nearby 3:2 mean-motion resonance also lacks a cold component
\citep{2001AJ....121.2804B,2008ssbn.book...59K}, which argues that if it swept
slowly through the 36--39~AU region the cold component must have already been
removed; otherwise, \citet{2005AJ....130.2392H} show that the low-$i$ objects
should have been captured into the 3:2 and preserved (because the 3:2 shields
its members from the effects of the $\nu_8$).  One possible interpretation is
that the 3:2 only obtained particles from a scattering hot population (as in
the \citet{2008Icar..196..258L} model) and ended with a large jump to its
current location, but a reason for the lack of a cold population inside 39~AU
would need to be provided.  Because the 3:2 location depends on only the
semimajor axis of Neptune, one might expect that the 3:2 resonance's arrival at
its current value would occur before the $\nu_8$ reaches its current location
due to the latter's dependence on the orbital elements of multiple planets and
the existence of other remaining mass in the system \citep{2000AJ....120.3311N}.

\subsection{The outer edge of the hot belt}
\label{sec:outer}

We successfully construct a model of the non-resonant, non-scattering TNOs with
semimajor axis beyond the 2:1 resonance by simply extending the L7 main-belt model out
into this region.
Using the classification system from \citet{2008ssbn.book...43G}, our current
sample contains 3 outer-belt TNOs and 11 detached TNOs; the
distinction between them is set by an arbitrary cut in eccentricity at
$e=0.24$.  For our current analysis, we group these two populations, under the
hypothesis that they share a smoothly-varying orbital
distribution.\footnote{Although it remains to be seen if the very
  large-inclination objects like Buffy \citep{2006ApJ...640L..83A} or Drac
  \citep{2009ApJ...697L..91G} are part of such a distribution.}. In order to
avoid problems with the exact border of the 2:1, we start our modelling at
$a=48$~AU; this eliminated 1 detached TNO, reducing our sample
to 13.

The
outer/detached objects share the same $(q,i)$ distributions as the hot main
classical belt. This suggests that again (as for the inner belt) the outer
population may be a smooth extension of the main-belt hot component.  To model
the outer/detached TNOs, we thus use the same prescription as for the
hot-main classical belt, with $\alpha=0.8$ and an $a$ range from 48~AU to a
value $a_{\rm max}$, with density varying as $a^{-\beta}$, with $\beta=2.5$.
We tried varying the exponent $\beta$ of the $a$ distribution.  For shallow
distributions, i.e. $\beta \le 1.5$, the model is rejected when $a_{\rm max}$
exceeds $\sim$100~AU, because it creates too many simulated detections
close to $a_{\rm max}$.  
The range $2.0 \le \beta \le 3.0$ produces acceptable models with no constraint on
$a_{\rm max}$.  Models with larger values of $\beta$ exhibit a very steep
decrease of number density at large $a$ and fail to produce enough
detections with $a>$60~AU.  We thus adopt $\beta=2.5$, as for
the main classical belt.  The number of objects needed to reproduce our 13
outer/detached detections is insensitive to our choice of $a_{\rm max}$ due to
the strong detection biases.  Hence we formulate our population estimate for a
population with no outer edge, finding a population beyond 48~AU of $N_{\rm
  outer/detached} ({\rm H}_g \le 8.0) = 10,000^{+7,000}_{-5,000}$ (see
Table~\ref{tab:pops}). 
Of these, only a small number 
$N_{\rm outer} ({\rm H}_g \le 8.0) = 500^{+350}_{-250}$ 
have $e < 0.24$, thus belonging to the outer belt defined
by \citet{2008ssbn.book...43G}.

As for our analysis of the inner belt, we computed the number density of TNOs
per unit $a$ at a main/outer interface at 47~AU, 
$P_{\rm outer/detached} ({\rm H}_g \le 8.0, 47\;{\rm AU}) = 340^{+230}_{-150}\, {\rm
  AU}^{-1}$ 
and compare it to the value from the outer edge of the main belt 
$P_{\rm main,outer} ({\rm H}_g \le 8.0) = 490^{+110}_{-100}\, {\rm AU}^{-1}$.  
Hence the TNO number density per unit $a$
in the outer/detached belt is the same as that of the hot main belt, within
uncertainties.  
There was absolutely no coupling in the debiasing procedure of these two 
TNO populations; this matching result was not tuned in any way.
Figure~\ref{fig:innerHotOuter} demonstrates that an initially-uniform
semimajor axis distribution for all three of these Kuiper belt sub-components 
as a single dynamical population is a plausible scenario.

Given the number of papers discussing a noticable edge to the distribution
\citep{1998AJ....115.2125J,2001ApJ...549L.241A,2001ApJ...554L..95T} this
continuity may be surprising.  Realize that the continuity is in the hot
component, which our analysis indicates is actually present throughout the
region from Neptune to at least several hundred AU.  This population has a
pericenter distribution with very few $q$'s above 40~AU, and may very well have
been emplaced as a sort of {\it fossilized scattered disk}
\citep{2002Icar..157..269G} as illustrated in \citet{2004AJ....128.2564M}.
This same process, however, does not emplace the kernel and stirred components
which dominate the main-belt region for $H_g > 8.0$, nor produce the dramatic
fall-off beyond $45$~AU in these cold populations.

\section{The scattering disk}
\label{sec:scat}

If the Centaurs and then JFCs do indeed come from one of the Kuiper Belt's 
sub-populations,
then their penultimate meta-stable source will be the set of TNOs currently 
scattering off Neptune, as defined by
\citet{2004MNRAS.355..935M} and \citet{2008ssbn.book...43G}.  Hence we wish to
give a population estimate for this `actively scattering' population.
Unfortunately the region occupied by the scattering objects is not a
simply-connected region definable by a simple parameter-space cut; they are
intimately mixed with stable resonant and non-resonant objects and providing a
full dynamical model of this region is well beyond the scope of the current
manuscript.  Here we examine available models of the scattering disk using the
CFEPS Survey Simulator to provide an order of magnitude population estimate for
this important transient population.

The definition of the {\it scattering} population has evolved over the last 15
years.  A cosmogonic perspective is easily adopted by workers doing
numerical simulations. In such simulations the `scattered' disk is taken to be
comprised of TNOs that currently do not have encounters with Neptune but were
delivered onto those orbits via an encounter.
\citet{2004MNRAS.355..935M} quantified this definition by requiring that
scattered TNO needs to have it semi-major axis change by more than 1.5~AU over
the life of the Solar System.
In this process, knowledge of orbital history is required for classification
and, clearly, this information is not available for a given real TNO.
More problematically, if the \citet{2008Icar..196..258L} model is correct then
the entire Kuiper Belt would qualify as having scattered off Neptune, making
the term {\it scattered disk object} a meaningless distinction.
\citet{2008ssbn.book...43G} proposed a practical definition for classification
based on the orbit of known objects at the current epoch, in which the
`scattering objects' are those {\it currently} (in the next 10~Myr) undergoing 
scattering encounters with Neptune in a forward simulation.
In the current manuscript we consider two definitions of the scattering disk,
one based on a parameterized region of phase space and one based on numerical
modelling of a particular scattering process, to derive an estimate of the
scattering population.

\citet{2000ApJ...529L.103T} and \citet{2005AJ....130.2392H} both give
population estimates of the scattered Kuiper Belt, but based on different
definitions of this population. The former called the scattered Kuiper Belt the
region of phase space $50~\AU \le a \le 200~\AU$ and $34~\AU~\le q
\le~36~\AU$. Based on their detection of 4 objects with preliminary
orbits in this region, they provide a population estimate of 18,000--50,000
objects (1-$\sigma$ range) with $D>100$~km, assuming an H$_g$
distribution slope of 0.8.  Using CFEPS and the same orbit and H$_g$
distributions as \citet{2000ApJ...529L.103T} we estimate the population in that
region of the phase space to be 2,100--17,500 objects (95\% confidence range),
about a factor of 4 less than \citet{2000ApJ...529L.103T}'s estimate.  This
estimate is based on scaling the Survey Simulator's detections to match
all the L7 detections in this range of $q$. 
Awkwardly, {\it none} of the 
L7 detections with orbits in this $q$ range are 
actually members of the scattering class, thus
this estimate is more correctly an estimate of some restricted portion of the
Detached population.  
Two of the objects (1999 CV$_{118}$ and 1999 CF$_{119}$)
used by \citet{2000ApJ...529L.103T} for their population estimate were later
found to not have orbits in the region they termed the scattered disk, hence
their population estimate of this region should be divided by 2, making it
more compatible with our estimate\footnote{The other two sources (1999
  TL$_{66}$ and   1999 CY$_{118}$) are found to be on scattering obits
  \citep{2008ssbn.book...43G}.}. 
The choice of the $q$=34--36~AU region was motivated by the candidate
`scattering' orbits known to \citet{2000ApJ...529L.103T}, intending this to be
a source region for the Centaurs and JFCs, as postulated by
\citet{1997Sci...276.1670D}.  However, the majority of the known TNOs in that
phase space cut are not currently interacting with Neptune and are on resonant
or detached orbits \citep{2008ssbn.book...43G}. A simple phase-space cut is
not appropriate for the scattering disk population.

To obtain an order-of-magnitude population estimate via a
dynamical model, we used the result of numerical integrations by
\citet{2006ApJ...643L.135G}.  This model attempted to produce the detached
population via secular interaction with rogue planets, where the additional
planet persists for the first 200 Myr of the simulation.
\citet{2006ApJ...643L.135G} find that the scattering particles in their
simulations that survive to the end of the a 4 Gyr integration largely forget
their initial state.  To obtain a scattering disk model, we selected the
orbital elements of actively scattering test particles during the last 500~Myr
of one 4.5-Gyr integration.
We then slightly smeared the orbital elements and applied an $H$-magnitude
distribution with slope $\alpha$=0.8.  We found a reasonable match between the
orbital elements of the L7 scattering sub-population and our input scattering
model, as observed by the Survey Simulator, althought the inclination
distribution was somewhat too cold, yielding a confidence level of only 8\%.  
The apparent magnitude distribution was rejected at more than 99\% whatever the
slope of the $H_g$ distribution we used; it is plausible that this rejection
is due to a change in luminosity-function slope in the size range
probed by our observations, as the faintest absolute magnitude of our detections
is $H_g = 10$ because scattering TNOs include many $q<30$~AU members. 
Three of the CFEPS active scatterers were inside 30 AU at the time of 
detection.  
The match between the orbital model and the observations allow us to be 
reasonably confident our population estimate is good to a factor of ten 
and we do not feel this order of magnitude estimate warrants further tuning 
until a larger sample of scattering objects is in hand.  
While there is clearly future room to better test models,
we give here the first published estimate of the active scattering population.  
The results are given
in Table~\ref{tab:popscat} for $H_g<10$ and for diameter $D>$100~km
($H_g<9.16$ assuming an albedo of $p_g$=0.05).  The quoted factor-of-three
uncertainty accounts only for the Poisson variation.

\begin{deluxetable}{lcc}
\tabletypesize{\footnotesize}
\tablecolumns{3}
\tablewidth{0pc}
\tablecaption{Scattering disk population estimates.
\label{tab:popscat}}
\tablehead{
\colhead{\ } & \colhead{$N({\rm H}_g \le 10)$} &
\colhead{$N({\rm D} > 100{\rm km})$} 
}
\startdata
Scattering disk & $25,000^{+20,000}_{-15,000}$ & $5,000^{+5,000}_{-3,000}$ \\
\enddata
\end{deluxetable}

We estimate an actively-scattering population that is about 2-3\% that of the
sum of the classical belts.  Interpretation of this number is problematic.  A
very large actively-scattering population would require that the current disk
could not be the steady state intermediary between the Centaurs and a
longer-lived source in the trans-neptunian region, in which case the
currently actively-scattering population is more likely to be the long-lived
tail of a roughly 100$\times$ more populous primordial population
\citep{1997Sci...276.1670D}.  For the 2-3\% figure, the active scatterers could
conceivably be now dominated by objects that have left the resonant, detached,
or classical populations in the last Gyr.

\section{Testing cosmogonic Kuiper belt models}
\label{sec:models}

The CFEPS-L7 model is an empirical parametric model that properly reproduces
the observed orbital distribution of the Kuiper belt, once passed through our
survey simulator.  The purpose of this parametric model is to provide
absolutely-calibrated population estimates of the various sub-populations of
the Kuiper belt.
The model also exhibits important features of the intrinsic Kuiper belt that a
cosmogonic model should reproduce.  For example, one needs to produce a cluster
of objects at low inclination and low eccentricity near 44~AU, that we call the
kernel.  There is also a low-$i$ component extending from the outer edge of the
$\nu_8$ secular resonance at 42.4~AU out to the 2:1 MMR with Neptune.  Finally,
there is a hot component with a confined $q$ range that extends in
semimajor-axis from the inner belt at $\sim$35~AU out to several hundred AU
with a decreasing surface density.  The synthetic L7 model is also useful for
observational modeling of our Kuiper belt, with \citet{2010AAS...21534606S} as
an example for the outer Solar System dust distribution based on the L3 model.

The ability to provide a detailed quantitive comparison with a cosmogonic model
is, however, the true power of the CFEPS survey.
This is done by passing a proposed model of the current Kuiper Belt
distribution through the CFEPS survey simulator and then comparing this
detection-biased model with the real CFEPS detections.  Through this procedure
one can choose between models in a statistically robust way.
Both the CFEPS L7 synthetic model and the CFEPS survey simulator are available
from the project web site {\it www.cfeps.net}.

Several models have been proposed to explain the dynamical structure of the
Kuiper belt \citep[][to name a few]{1993Natur.365..819M,
  2000ApJ...528..351I,2005AJ....130.2392H,2008Icar..196..258L}.  Since the
primary purpose of CFEPS was to validate or refute cosmogonic models, we
present an example of this process.  Because we had available both an orbital
element distribution and a resonance-occupation analysis (Levison, 2010,
private communication), we have chosen to use Run B of
\citet{2008Icar..196..258L} as an example of how one uses the CFEPS Survey
Simulator to compare a model to the observed Kuiper belt.  The simulation in
question (motivated by the Nice model of the re-arrangment of the outer Solar
System) has already some known problems pointed out by its authors, but the
model's intriguing aspects make it a good example of the comparison process.

All the fictitious TNOs in the Run B model were dynamically classified
following the \citet{2008ssbn.book...43G} procedure (Van Laerhoven, 2010,
private communication).  The final planetary configuration in the Nice model
was intentionally made different from that of the Solar System to avoid the
$\nu_8$ secular resonance inadvertently sweeping through and destroying the
belt and thus has many objects in the 40~$< a <$~42.4~AU range at low
inclination, where the real $\nu_8$ resonance would eliminate them.  To avoid
this complication, we restricted the comparison to the range 42.4~$< a
<$~47~AU, yielding 128 non-resonant model TNOs from run B.

We use the following procedure to generate the large number of TNOs required as
input to the survey simulator. First we select a model object at random, and
vary the orbital elements uniformly by $\pm0.2$~AU in both $a$ and $q$ and
$\pm0.5^\circ$ in inclination and then randomize the elements $\Omega$,
$\omega$, and $M$, There was no size distribution given for the model objects,
so the $H_g$ magnitude of each object was drawn from an exponential
distribution (eq.~\ref{eq:hdistrib}).  Given the orbital elements and $H_g$
magnitude, we then use our Survey Simulator to determine if the model object
would have been detected by the CFEPS-L7 obersvations.  We repeat the procedure
until we have a set number of simulated detections and then compare the $a$,
$e$, $i$, $q$ and $r$ one-dimensional cumulative distributions of the simulated
detections to those of the L7 sample, using the Anderson-Darling test.

For the $H_g$ distribution, we tried single slopes of $0.6 \le \alpha \le 1.3$
and also a model with $\alpha=1.2$ for the low-$i$ and $\alpha=0.8$ for the
high-inclination objects, as in our favored model.
All the models for the $H_g$ distribution produced acceptable matches to the
apparent magnitude distribution, but had no effect on the orbital element
distributions, so we do not show the magnitude distribution.

\begin{figure*}[h]
\begin{center}
\includegraphics[width=18cm]{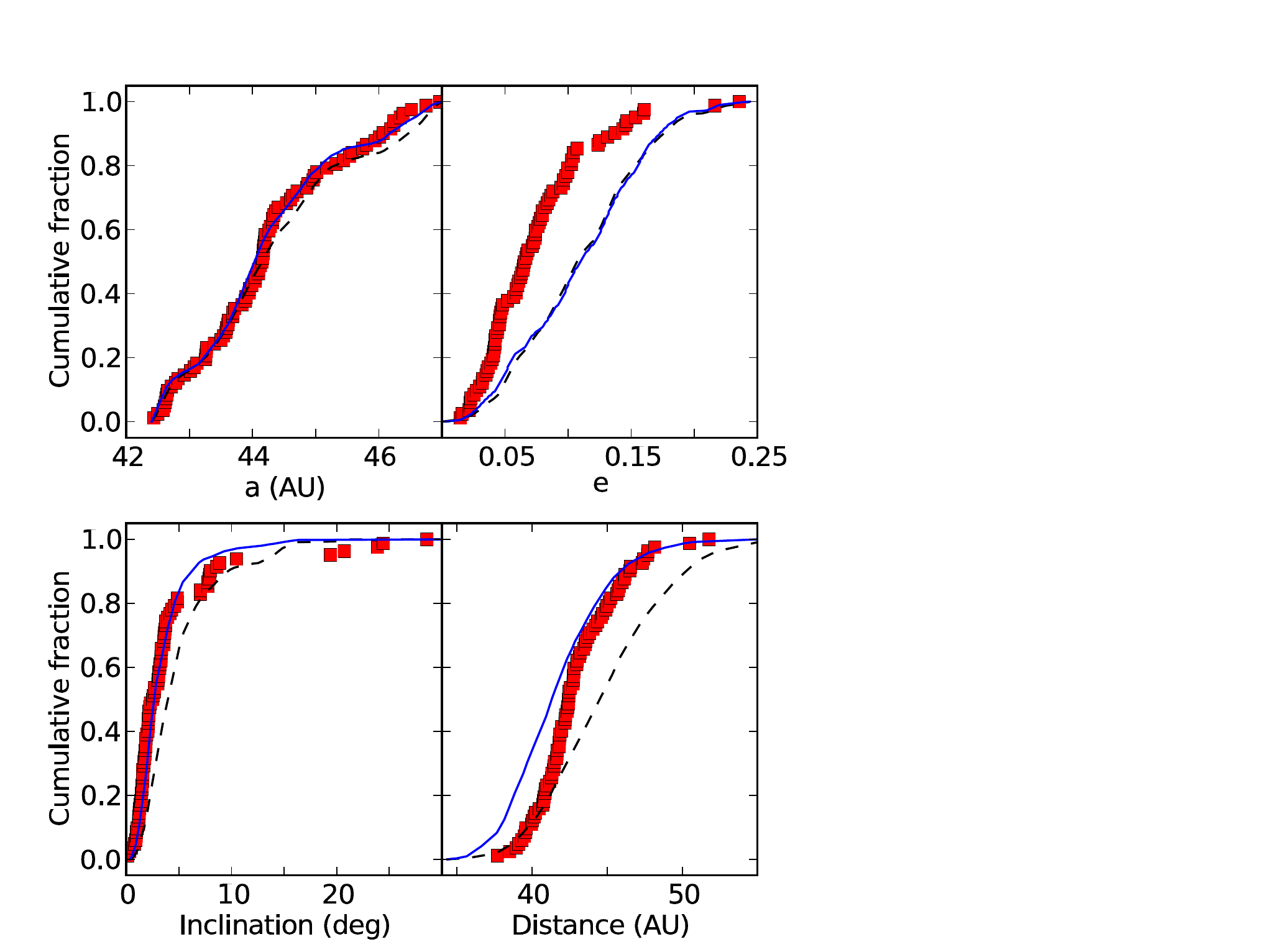}
\end{center}
\caption{CFEPS+Pre objects (red solid squares) compared to the Nice model's
  distribution in $a$ (upper left), $e$ (upper right), $i$ (lower left) and
  $r$ (lower right) distributions when the intrinsic (black dashed line)
  distribution is biased via the CFEPS Survey Simulator (resulting blue solid
  lines). This shown case corresponds to a single-slope $\alpha=1.1$ $H_g$
  distribution.  }
\label{fig:cdfnice}
\end{figure*}

Fig.~\ref{fig:cdfnice} compares the distribution of $a$, $e$, $i$ and $r$ of
the L7 sample and simulated detections from the model.  There is
remarkably-good agreement for the $a$ distribution.  Although the Nice model
does not exhibit a clustering around 44~AU as strong as the L7 sample,
the difference between the two distribution is not statistically significant.

On the other hand, the model's $e$ distribution is too excited compared to the
observed one, as already noted by \citet{2008Icar..196..258L}.  This then
results in detection distances that are overly dominated by small-distance
detections.  For both cumulative distributions, the Anderson-Darling test says
that the hypothesis that the observed objects could be drawn from the model can
be rejected at $>$99.9\% confidence.

The model's $i$ distribution is not a good match either, again as already noted
by \citet{2008Icar..196..258L}; the AD test rejects the $i$ distribution at
more than 99.9\% confidence.  This is mostly because the L7 distribution has
two components, while the Run B input model gives an essentially unimodal
distribution.  The simulated detections from the model appear roughly
consistent up to $i\simeq 4^\circ$, but there is a lack of high-$i$ TNOs, to
which the Anderson-Darling test is sensitive.  Run B lacks the hot component
that peaks between 15$^\circ$ and 20$^\circ$ and extends past 30$^\circ$.
Looking only at the $i<6^\circ$ region, here too the AD test rejects the model
at more than 99.9\% confidence.  Most of the Run B classical belt comes from
the low-$i$ outer part of the planetesimal disk, which acquires inclinations
similar to that of the inner initial disk, with a width $6^\circ$.  The true
cold component, on the contrary, has a width 
certainly less than $3^\circ$.
We conclude
that one needs a strongly bi-modal input population to the Nice model in order
to produce the desired bi-modal inclination distribution we see in the real
Kuiper belt.  More worrying is the claim by \citet{2008Icar..196..258L} that
increasing the width of the initial population produced the same final
inclination distribution, meaning there is a missing ingredient in this model
to explain the dynamical structure of the Kuiper belt.
\citet{2008Icar..196..258L} mention another simulation, Run E, which generated
too many hot objects compared to the cold population; so there may be an
intermediate parameter set that could match the observations.

\section{Conclusions}
\label{sec:concl}

This paper's modeling concerns the non-resonant Kuiper Belt,
although the L7 release lists all detections from the CFEPS survey fields from
2002--2007 for the sake of completeness. 
Due to the complexity of the modelling required because of the
phase relations with Neptune, the resonant populations are presented 
in a separate paper \citep{2011AJ......1....1G}.
We find that the debiased orbital and $H$-magnitude distributions show 
that there is considerable sub-structure in the main Kuiper Belt. 
We quantitatively measured the size of the various sub-populations, 
create an empirical model of these sub-populations in the L7 synthetic
model, and provide an algorithm (the CFEPS survey simulator) to
quantitatively compare cosmogonic models to the intrinsic Kuiper Belt.  
Here we
summarize the results and offer a synthesis and interpretation.

A plausible hypothesis is that the hot population permeates the entire Kuiper
Belt region from 30~AU up to at least 200~AU, albeit with a projected
surface density (onto the invariable plane) that decreases with semimajor 
axis.
Even the resonant populations are consistent with the idea that the entire hot
component is a vestigal ``fossilized'' scattered disk from an epoch when TNOs
with perihelia up to $\sim$40~AU were being weakly scattered by a massive 
object at the inner edge of the Kuiper Belt (whether this object was Neptune 
or something else is unclear from the present data).  
The inclination distribution of this hot population can be represented by
$\sin{(i)}$ times a gaussian of width $\sim 16^\circ$.  
Note however that, due to the strong bias against
detection of large-$i$ objects in an ecliptic survey, our current sample does
not provide a strong constraint on the width or the functional form of the
hot component.  A scenario in which the inner belt, hot main-belt, outer belt
and detached populations, along with the resonant populations were all emplaced
simultaneously from a population scattered outward during the final stages of
planet formation, with a single size distribution, initial inclination
distribution, colour distribution and binary fraction, is an attractive
hypothesis.  The plausibly-continuous initial number density across the
inner/main and main/outer boundary (see Fig.~\ref{fig:innerHotOuter})
supports this idea.

The Kuiper Belt's (surviving) `cold' population is entirely
confined between semimajor axes of 42.4~AU and the 2:1 resonance with Neptune, 
and its inclination distribution (measured relative to the J2000 ecliptic)
is adequately represented via $\sin{(i)}$ times a gaussian of 
width 2.6$^\circ$, with an acceptable range from 2.3$^\circ$ up to 
3.5$^\circ$.
There are indeed $i<5^\circ$ TNOs in the low-$i$ tail of the hot population all
over the Kuiper belt and even in the $42.4<a<47$~AU region, so an inclination
cut does not provide a clean separation between the hot and cold components of
the main belt. In the current belt we claim that all $i<5^\circ$ TNOs with
semi-major axis outside the above range are the hot-component objects that
happen to have lower inclinations.  The cold population exhibits a
particularly-strong grouping in band of about 1-AU $a$ thickness, centered at
44~AU (which we call the kernel).  The linear number density (\#/AU) 
of `cold' belt objects increases from the inner edge at 42.4~AU up to a 
maximum at $\sim 44.4$~AU, all with rather low eccentricities.  
Past 44.4~AU, the linear number density drops
noticeably, and classical TNOs tend to have higher eccentricities; 
the CFEPS-L7 model uses a `stirred' population that covers the 
42.4--47~AU range with a single parameterization.  
We favor the idea that this cold component is primordial
(the objects formed at roughly their current heliocentric distances),
although this is not required.

The primordial distance range of the cold population is difficult to 
constrain.
The inner boundary at $a$=42.4~AU may have been eroded via
scattering by massive bodies and resonance migration; an important condition is
that any sequence of events cannot allow either the inner belt, or the
mean-motion and secular resonances that probably migrated through it, to have
preserved a cold component today.  
The coincidence of the stirred population's outer edge with the 2:1
resonance suggests to us that the kernel marks the original outer
edge and that the larger-$a$ cold objects have either (i) been dragged out of
the $a<44.4$~AU region via trapping and then drop-off in the 2:1 as it went
past (in the fashion studied by \citet{2005AJ....130.2392H}) or (ii) due to
weak scattering out of the $40<a<44.4$~AU region.
Perhaps
the edge of the original cold population around 45~AU may be explained
by the global evolution of solid matter in turbulent
protoplanetary disks \citep{1996A&A...309..301S,1997A&A...319.1007S},
although an even-more extreme density contrast may be needed at $\sim$30~AU
to prevent Neptune's continued migration outward \citep{2004Icar..170..492G}.
Sharp drops in surface density are commonly observed in protoplanetary
disks at about this 30--50 AU scale
\citep{1998ApJ...499..758J,2009ApJ...694L..36M,2010ApJ...725..430M}.

There is an issue with a primordial origin of the cold population at this
location. The on-ecliptic mass density of this population is extremely low and
it would be difficult to form multi-hundred km TNOs in a low surface density
environment.  This may not be impossible due to recent work on forming
planetesimals big \citep{2009Icar..204..558M,2011arXiv1102.4620Y}, which can be
favored by external photoevaporation \citep{2005ApJ...623L.149T}, and may be
supported by the fact that it appears that there are simply no cold objects
larger than $H\sim5$; all the larger objects are in the other populations which
may come from closer to the Sun where the mass density was higher.

The kernel around 44~AU is an intriguing feature.  A collisional family
explanation would eliminate the idea that the 44.4-AU edge is a primordial
edge, but would instead simply be where the very-low velocity dispersion 
breakup occured.
This velocity dispersion is even lower than for the putative Haumea collisional
family \citep{2007Natur.446..294B}.  
An additional puzzle is the unclear significance
that the kernel is bounded between the 7:4 and the 9:5 MMRs .  One
possible, very ad-hoc, explanation would be that the 2:1 MMR started its
migration interior to 43.5~AU while being wide (due to a large Neptune
eccentricity), and then had a stochastic jump by a few tenth of AU while near
44~AU, leaving behind a pile of objects that we see as the kernel today.

The hot population poses other strong constraints.  
Models by \citet{2003Icar..161..404G,2005AJ....130.2392H,2008Icar..196..258L}
all succeed in creating a hot population that has a similar radial extent to
what is currently observed, but have varying success in matching the
inclination distribution.  
When slowly migrating Neptune over long distances ($>8$~AU) into an
initially-cold disk, \citet{2003Icar..161..404G} and
\citet{2005AJ....130.2392H} generated a reasonable TNO fraction with
inclinations up to 35$^\circ$.
When migrating Neptune over a shorter distance 
\citep{2003Icar..161..404G}, or in a hot disk
\citep{2005AJ....130.2392H}, the fraction of high-$i$ TNOs is noticeably
reduced, while still reaching the same maximum $i$.
\citet{2008Icar..196..258L}, on the contrary, migrate Neptune over a short
distance (2--3~AU) into a warm scattered disk (with $\langle i \rangle =
6^\circ$) and essentially maintain the input inclination distribution.  They
report that increasing the initial $i$ distribution resulted in the same final
population, which lacks TNOs with $i>30^\circ$ and which we have confirmed is
colder than the actual belt.  These facts appear to indicates that Neptune had
to slowly migrate over a long distance in a cold disk in order to obtain the
observed inclination distribution of the hot population.  However,
\citet{2009A&A...507.1041M} showed that a long and slow migration of Neptune,
coupled with a similar migration of the other giant planets, does not correctly
reproduce the secular architecture of the solar system, in particular the
amplitudes of the eigenmodes characterising the current secular evolution of
the eccentricities of Jupiter and Saturn.  They conclude that only the Nice
model can reproduce the current dynamics of the inner solar system and the
giant planets.  Unfortunately this scenario does not produce Kuiper-Belt
components with orbital properties that agree with the L7 orbit catalog.

The idea that the hot population originated from a planetesimal population
scattered outward by Neptune, whose resonant and largest-$q$ members are
preserved, is extremely attractive.  Thus much of \citet{2008Icar..196..258L}'s
general scenario has many pleasing aspects and one is tempted to think of the
hot population as the transplanted population, even if our results show that
the inclination distribution is a stumbling block.  
Contrary to some statements \citep[$eg$][]{2010Icar..210..944F}, 
we find that the Nice model is not good at producing the 
the hot population's inclination distribution, 
but suprisingly produces rather well the cold population's
fine structure in the semimajor axis distribution.  
The large-$i$ TNOs which do appear could instead be interpreted as coming
from the `evader' mechanism of \citet{2003Icar..161..404G}.  
In this conception the \citet{2010Icar..210..944F} finding, that the 
luminosity distribution of the Jovian Trojans is more similar to the 
cold than hot TNO populations, makes perfect sense in a scenario in which the 
injection of bodies into the Jovian Trojan region occurs 
from the same source region as 
the implantation of the Kuiper Belt's cold population. 
In the Nice model this seems unlikely because jovian Trojan
capture occurs just after the Jupiter-Saturn mutual 1:2 resonance crossing 
\citep{2005Natur.435..462M} 
and
involves small bodies closer to the planets than the cold outer disk that is
the main source of the cold Kuiper belt.  
If so, the hot component cannot be
generated from the Nice model's inclined inner disk, as this would have the
same size distribution as the Jovian Trojans.  One needs another source for the
hot population, one that is not too perturbed by the initial instability in
Neptune's motion.  A final caveat concerns the existence of wide binaries in
the cold belt; 
\citet{2010ApJ...722L.204P} showed that the Neptune scattering occuring in 
the Nice model would disrupt nearly all 
such wide binaries, thus requiring a more gentle mechanism to move 
the cold-belt to its current location if that population did not 
form {\it in-situ}.

Our current understanding of the trans-neptunian region is not likely to
advance rapidly for time scales of order a decade unless new surveys begin to
efficiently probe TNOs that were rare in the ecliptic surveys.  The most likely
approach that would result in an advance are moderate-depth (24th magnitude)
wide-field surveys (many hundreds of square degrees) at higher ecliptic
latitudes, or deeper (25th magnitude) surveys covering
$\sim100$~sq.~deg. targeting regions of sky that attempt to isolate
cosmogonically-interesting sub-populations.  
We hope that CFEPS will serve as a standard for the need for 
well-characterized discovery and tracking.  
The Large Synoptic Survey Telescope (LSST) \citep{2008SerAJ.176....1I} 
should certainly firm up the main-belt dynamical sub-structure along with 
the colour and size distributions for those components.

\appendix
\section{Appendix A}
\label{sec:app_a}

In this Appendix, we give details of the algorithm used to generate the
CFEPS-L7 model of the main classical belt.

The main classical belt objects are constrained in what is essential
3-dimensional phase-space due to the (confirmed a posteriori) fact that the
mean anomaly and longitudes of ascending node and perihelion are all uniformly
distributed in the intrinsic population.  Thus the L7 model consists of 3
sub-populations constrained by 3 orbital-element distributions to determine for
each sub-population.

The inclination distribution of each subcomponent is well represented by a
probability distribution proportional to $\sin(i)$ times a gaussian
$\exp[i^2/(2\sigma^2)]$, where past results indicate a `cold'-component width of
  $\sim 2.5^\circ$ and a `hot'-component width of $\sim 15^\circ$
  \citep{2001AJ....121.2804B,2008ssbn.book...59K}.

The {\it hot} component occupies the semimajor axis range from 40.0 to 47.0
AU and is defined by:
\begin{itemize}
\item an $a$ distribution with a Probability Density Function (PDF)
  proportional to $a^{-5/2}$, corresponding to a surface
  density proportional to $a^{-7/2}$;
\item an inclination distribution proportional to
$\sin(i) \times \exp [ i^2/(2\sigma_h^2) ]$, with width $\sigma_h = 16^\circ$;
\item we eliminate objects from the region unstable due to
  the $\nu_8$ secular resonance: $a<42.4$~AU and $i<12^\circ$;
\item a perihelion distance $q$ distribution that is mostly uniform between
  35 and 40~AU, with soft shoulders at both ends extending over $\sim$1~AU; the
  PDF is proportional to 
   $1/([1+\exp{((35 - q)/0.5)}] [1+\exp{((q - 40)/0.5)}])$; 
  any object with $q<$34~AU is rejected;
\item finally, we reject objects with $q < 38 - 0.2 i$~(deg) to account for 
  weaker stability of low-$q$ orbits at low inclination.
\end{itemize}
We have found that the exact form of the truncation at low
perihelion distance is unimportant, as long as the limiting value of $q$ is a
decreasing function of the inclination; this is justified dynamically as
low-inclination orbits cannot have $q < 38$~AU and remain stable
\citep{1995AJ....110.3073D}.

The {\it stirred} component covers only the range of stable semimajor axis at
low inclinations:
\begin{itemize}
\item an $a$ distribution with PDF
  proportional to $a^{-5/2}$ between 42.4 (limit of the $\nu_8$
  resonance) and 47~AU;
\item a uniform $e$ distribution between 0.01 and a maximum value depending
  on the semimajor axis, $e_{max} = 0.04 + (a - 42)\times0.032$, to 
  reproduce the structure seen in  figs.~\ref{fig:CFEPS} and \ref{fig:MPC};
\item randomly keep objects with probability
  $1/(1+\exp{[(e-0.6+19.2/a)/0.01]})$, which corresponds to a soft cut at $q =
  38 + 0.4*(a - 47)$;
\item an inclination distribution proportional to $\sin(i)$ times
  a gaussian of width $\sigma_c = 2.6^\circ$;
\item again, we reject objects with $q < 38 - 0.2\times i$(deg) as for the hot
  component.
\end{itemize}

Finally, the {\it kernel} provides the group of objects with low inclination in
the middle of the main classical belt as seen in fig.~\ref{fig:CFEPS}:
\begin{itemize}
\item a uniform $a$ distribution between 43.8 and 44.4~AU;
\item a uniform $e$ distribution between 0.03 and 0.08;
\item an inclination distribution proportional to $\sin (i)$ times
  a gaussian of width $\sigma_c = 2.6^\circ$, identical to the
  stirred population.
\end{itemize}

For all components, the remaining orbital elements (longitude of node, argument
of perihelion and mean anomaly) are drawn at random uniformly between 0$^\circ$
and 360$^\circ$.
All elements are generated in the invariable plane reference frame (inclination
1$^\circ$~35'~13.86" with respect to J2000 ecliptic plane with direction of
ascending node at 107$^\circ$~36'~30.8"). In particular, we state widths of the
inclination distribution with respect to the invariable plane.
\citet{2005AJ....129.1117E}, \citet{2004AJ....127.2418B} and
\citet{2010AJ....140..350G} studied the distribution of inclinations with
respect to their self-determined Kuiper Belt Plane, which differ from the
invariable plane.

To evaluate the acceptability of each model we evaluate our parameterization in
distinct portions of phase-space.
\begin{itemize}
\item $i \ge 10$~deg
\item $i < 10$~deg
\item $a > 44.4$~AU
\item $a \le 44.4$~AU
\item the entire main-belt region
\end{itemize}
We computed the probability of the AD or KKS statistics in each region
separately and consider the minimum on all element distributions and all
sub-regions when determining if a particular parameterization is rejected.

The variable parameters are the $i$-width of the hot component $\sigma_h$, the
cold component's $i$ width $\sigma_c$, the $H$-magnitude distribution of these
two components (with slopes $\alpha_h, \alpha_c$), the hot population's
fraction of the main belt $f_h$, and the kernel fraction $f_k$, with the
stirred component forming the remainder: $f_s = 1 - f_h - f_k$.

The CFEPS-L7 model has the following known weaknesses:
\begin{enumerate}
\item Resonant orbits will be generated by chance in the main-belt region
(especially for the 5:3, 7:4, and 9:5 resonances).
\item The $\nu_8$ resonance cut is done in osculating, rather than proper, 
orbital elements space and thus some L7 model objects objects near the
resonance will be unstable.
\item There are four tiny semimajor axis gaps in our model: small regions 
   ($\sim0.3$ AU in $a$) on both sides of the  3:2 and 2:1 resonannces.
\end{enumerate}

\section{Appendix B}
\label{sec:app_b}

The CFEPS project is built on the observations acquired as the `Very Wide' component of the CFHT Legacy Survey (CFHTLS-VW).  
All discovery imaging data is publicly available from the Canadian Astronomy Data Centre (CADC\footnote{http://www.cadc.hia.nrc.gc.ca}). 
These images were acquired using the CFHT Queue Service Observing (QSO) system. 
For each field observed on a photometric night the CFHT QSO provides calibrated images using their ELIXIR processing  software \citep{2004PASP..116..449M}. 
Our photometry below is reported in the Sloan system \citep{1996AJ....111.1748F} 
with the calibrations contained in the header of each image as provided by ELIXIR.  Color corrections were computed using the average color for Kuiper belt objects $(g-r)\sim 0.7$.
Differential aperture photometry was determined for each of our detected objects observed on photometric nights and these fluxes are reported in Table~\ref{tab:phot}.  
All CFEPS discovery observations were acquired in photometric conditions in a relatively
narrow range of seeing conditions due to queue-mode acquisition.  
The photometry below supercedes information that may be in the Minor Planet Center's 
observational database.

{\small{
\begin{deluxetable}{lrrrrrrrrr}
\tabletypesize{\scriptsize}
\tablecolumns{7}
\tablewidth{0pc}
\tablecaption{Object Fluxes.\label{tab:phot}}
\tablehead{\colhead{Object} & 
\colhead{g} & \colhead{$\sigma_g$} & \colhead{N$_g$} & 
\colhead{r} & \colhead{$\sigma_r$} & \colhead{N$_r$} & 
\colhead{i} & \colhead{$\sigma_i$} & \colhead{N$_i$}}
\startdata

%\loaddata{tables/L3PhotometryTable}
L3f01    &  23.66 &  0.41 &  4 &  23.12 &  0.15 &  3 & \nodata & \nodata & .. \\ 
L3f04PD  &  22.74 &  0.33 &  4 & \nodata & \nodata & .. & \nodata & \nodata & .. \\ 
L3h01    &  23.83 &  0.27 &  4 &  23.02 &  0.11 &  3 & \nodata & \nodata & .. \\ 
L3h04    &  24.32 &  0.17 &  4 &  23.82 &  0.33 &  7 & \nodata & \nodata & .. \\ 
L3h05    &  24.36 &  0.07 &  2 &  23.56 &  0.30 &  7 & \nodata & \nodata & .. \\ 
L3h08    & \nodata & \nodata & .. &  23.15 &  0.75 &  7 & \nodata & \nodata & .. \\ 
L3h09    &  22.73 &  0.04 &  4 &  22.29 &  0.23 & 10 & \nodata & \nodata & .. \\ 
L3h11    &  23.44 &  0.20 &  7 &  23.13 &  0.10 &  3 & \nodata & \nodata & .. \\ 
L3h13    &  23.73 &  0.10 &  4 &  23.29 &  0.27 &  7 & \nodata & \nodata & .. \\ 
L3h14    &  23.27 &  0.15 &  3 &  22.81 &  1.42 &  8 & \nodata & \nodata & .. \\ 
L3h18    &  23.42 &  0.09 &  3 &  22.53 &  0.17 &  8 & \nodata & \nodata & .. \\ 
L3h19    & \nodata & \nodata & .. &  23.67 &  0.26 &  9 & \nodata & \nodata & .. \\ 
L3h20    & \nodata & \nodata & .. &  23.15 &  0.30 &  7 & \nodata & \nodata & .. \\ 
L3q01    &  23.89 &  0.21 &  3 &  22.96 &  0.17 &  3 &  22.76 &  0.36 &  3 \\ 
L3q02PD  &  23.50 &  0.10 &  3 &  22.49 &  0.07 &  4 &  22.23 &  0.02 &  3 \\ 
L3q03    &  23.19 &  0.17 &  4 &  22.36 &  0.25 &  4 &  22.37 &  0.00 &  1 \\ 
L3q04PD  &  24.15 &  0.43 &  4 &  23.31 &  0.14 &  4 &  23.10 &  0.16 &  3 \\ 
L3q06PD  &  23.58 &  0.30 &  4 & \nodata & \nodata & .. & \nodata & \nodata & .. \\ 
L3q08PD  &  23.67 &  0.18 &  3 & \nodata & \nodata & .. & \nodata & \nodata & .. \\ 
L3q09PD  &  23.50 &  0.24 &  4 & \nodata & \nodata & .. & \nodata & \nodata & .. \\ 
L3s01    &  23.54 &  0.12 &  6 & \nodata & \nodata & .. &  22.54 &  0.12 &  2 \\ 
L3s02    &  23.81 &  0.28 &  6 &  23.40 &  0.23 &  4 &  23.18 &  0.11 &  2 \\ 
L3s03    &  22.90 &  0.20 &  5 & \nodata & \nodata & .. &  22.65 &  0.23 &  3 \\ 
L3s05    &  23.67 &  0.30 &  5 & \nodata & \nodata & .. &  22.88 &  0.13 &  2 \\ 
L3s06    &  22.82 &  0.03 &  5 & \nodata & \nodata & .. &  21.89 &  0.07 &  3 \\ 
L3w01    &  22.89 &  0.61 &  5 & \nodata & \nodata & .. & \nodata & \nodata & .. \\ 
L3w02    &  23.56 &  0.11 &  4 &  22.80 &  0.08 &  4 &  22.54 &  0.07 &  3 \\ 
L3w03    &  23.76 &  0.07 &  5 &  22.50 &  0.08 &  4 & \nodata & \nodata & .. \\ 
L3w04    &  22.44 &  0.03 &  5 &  21.65 &  0.04 &  4 &  21.53 &  0.02 &  3 \\ 
L3w05    &  24.20 &  0.31 &  4 &  23.70 &  0.17 &  3 &  23.77 &  0.07 &  3 \\ 
L3w06    &  23.65 &  0.25 &  4 &  23.13 &  0.19 &  4 & \nodata & \nodata & .. \\ 
L3w07    &  22.95 &  0.09 &  5 & \nodata & \nodata & .. &  22.46 &  0.09 &  3 \\ 
L3w08    &  23.96 &  0.13 &  4 & \nodata & \nodata & .. &  22.79 &  0.14 &  3 \\ 
L3w09    &  23.53 &  0.10 &  3 &  22.72 &  0.15 &  4 & \nodata & \nodata & .. \\ 
L3w10    &  23.95 &  0.20 &  5 &  23.04 &  0.13 &  4 &  22.00 &  0.63 &  3 \\ 
L3w11    &  24.03 &  0.12 &  4 &  23.49 &  0.15 &  4 &  23.34 &  0.13 &  3 \\ 
L3y01    &  24.04 &  0.26 &  4 &  22.62 &  0.55 &  3 & \nodata & \nodata & .. \\ 
L3y02    &  23.38 &  0.09 &  6 &  22.69 &  0.19 &  4 & \nodata & \nodata & .. \\ 
L3y03    &  23.41 &  0.09 &  4 &  22.79 &  0.11 &  4 & \nodata & \nodata & .. \\ 
L3y05    &  23.89 &  0.03 &  4 &  22.99 &  0.12 &  3 & \nodata & \nodata & .. \\ 
L3y06    &  23.37 &  0.18 &  3 & \nodata & \nodata & .. & \nodata & \nodata & .. \\ 
L3y07    &  23.42 &  0.09 &  4 &  22.87 &  0.23 &  5 & \nodata & \nodata & .. \\ 
L3y09    &  23.65 &  0.17 &  4 &  23.01 &  0.14 &  5 & \nodata & \nodata & .. \\ 
L3y11    &  23.82 &  0.32 &  4 &  23.51 &  0.21 &  4 & \nodata & \nodata & .. \\ 
L3y12PD  &  21.73 &  0.03 &  4 &  20.81 &  0.04 &  4 & \nodata & \nodata & .. \\ 
L3y14PD  &  23.68 &  0.19 &  4 &  22.73 &  0.10 &  3 & \nodata & \nodata & .. \\ 
l3f05    &  23.71 &  0.18 &  3 & \nodata & \nodata & .. & \nodata & \nodata & .. \\ 
l3h10    & \nodata & \nodata & .. &  23.03 &  0.28 &  7 & \nodata & \nodata & .. \\ 
l3h15    & \nodata & \nodata & .. &  23.74 &  0.27 &  7 & \nodata & \nodata & .. \\ 
l3h16    & \nodata & \nodata & .. &  23.53 &  0.64 &  7 & \nodata & \nodata & .. \\ 
l3q05    &  23.68 &  0.16 &  4 & \nodata & \nodata & .. & \nodata & \nodata & .. \\ 
l3q07    &  24.19 &  0.21 &  7 & \nodata & \nodata & .. & \nodata & \nodata & .. \\ 
l3w14    &  23.98 &  0.20 &  3 & \nodata & \nodata & .. & \nodata & \nodata & .. \\ 
l3w19    &  24.00 &  0.11 &  3 & \nodata & \nodata & .. & \nodata & \nodata & .. \\ 
U3f02    &  24.07 &  0.21 &  4 &  23.41 &  0.31 &  3 & \nodata & \nodata & .. \\ 
U3h06    & \nodata & \nodata & .. &  23.96 &  0.35 &  7 & \nodata & \nodata & .. \\ 
U3s04    &  24.10 &  0.45 &  6 & \nodata & \nodata & .. & \nodata & \nodata & .. \\ 
U3w13    &  24.39 &  0.21 &  4 &  23.96 &  0.21 &  3 & \nodata & \nodata & .. \\ 
U3w16    &  24.07 &  0.14 &  4 & \nodata & \nodata & .. &  23.14 &  0.25 &  3 \\ 
U3w17    &  24.45 &  0.09 &  4 & \nodata & \nodata & .. &  23.27 &  0.15 &  3 \\ 
U3y04    &  24.25 &  0.13 &  4 &  22.57 &  2.16 &  4 & \nodata & \nodata & .. \\ 
U3y08    &  24.25 &  0.13 &  4 &  23.43 &  0.25 &  3 & \nodata & \nodata & .. \\ 
U3y16    &  23.93 &  0.37 &  4 &  23.45 &  0.16 &  3 & \nodata & \nodata & .. \\ 
u3h02    & \nodata & \nodata & .. &  23.96 &  0.25 &  4 & \nodata & \nodata & .. \\ 
u3h03    & \nodata & \nodata & .. &  24.01 &  0.09 &  3 & \nodata & \nodata & .. \\ 
u3h07    & \nodata & \nodata & .. &  24.35 &  0.43 &  4 & \nodata & \nodata & .. \\ 
u3h12    & \nodata & \nodata & .. &  23.79 &  0.23 &  7 & \nodata & \nodata & .. \\ 
u3h17    & \nodata & \nodata & .. &  24.17 &  0.33 &  4 & \nodata & \nodata & .. \\ 
u3w12    &  24.43 &  0.15 &  3 & \nodata & \nodata & .. &  23.48 &  0.26 &  3 \\ 
u3w15    &  24.07 &  0.21 &  3 & \nodata & \nodata & .. & \nodata & \nodata & .. \\ 
u3w18    &  24.49 &  0.19 &  3 & \nodata & \nodata & .. & \nodata & \nodata & .. \\ 
u3y10    &  24.00 &  0.09 &  3 & \nodata & \nodata & .. & \nodata & \nodata & .. \\ 
u3y13    &  24.28 &  0.16 &  3 & \nodata & \nodata & .. & \nodata & \nodata & .. \\

%\loaddata{tables/L4PhotometryTable}
L4h01PD  &  23.77 &  0.40 &  3 & \nodata & \nodata & .. & \nodata & \nodata & .. \\ 
L4h02PD  &  23.50 &  0.18 &  3 & \nodata & \nodata & .. & \nodata & \nodata & .. \\ 
L4h05PD  &  23.96 &  0.16 &  3 &  23.48 &  0.31 &  4 & \nodata & \nodata & .. \\ 
L4h06    &  23.83 &  0.12 &  3 &  22.77 &  0.75 &  2 & \nodata & \nodata & .. \\ 
L4h07    &  23.70 &  0.08 &  3 &  23.35 &  0.37 &  6 & \nodata & \nodata & .. \\ 
L4h08    &  23.01 &  0.03 &  4 &  22.72 &  0.13 &  3 &  22.41 &  0.14 &  3 \\ 
L4h09PD  &  21.34 &  0.18 &  4 & \nodata & \nodata & .. & \nodata & \nodata & .. \\ 
L4h10PD  &  22.96 &  0.09 &  4 & \nodata & \nodata & .. &  22.62 &  0.16 &  3 \\ 
L4h11    &  22.98 &  0.09 &  4 &  22.52 &  0.10 &  4 &  22.08 &  0.09 &  3 \\ 
L4h12    &  24.11 &  0.34 &  4 &  22.92 &  0.35 &  4 &  22.60 &  0.01 &  3 \\ 
L4h13    &  23.64 &  0.23 &  3 &  23.34 &  0.13 &  3 &  22.53 &  0.15 &  2 \\ 
L4h14    &  23.68 &  0.30 &  4 &  23.65 &  0.31 &  3 & \nodata & \nodata & .. \\ 
L4h15    &  24.05 &  0.05 &  3 &  23.74 &  0.63 &  7 & \nodata & \nodata & .. \\ 
L4h16    &  24.04 &  0.45 &  3 &  23.38 &  0.25 &  3 & \nodata & \nodata & .. \\ 
L4h18    &  23.34 &  0.38 &  4 &  23.54 &  1.00 &  3 &  22.01 &  0.25 &  3 \\ 
L4h21    &  23.75 &  0.05 &  3 &  23.43 &  0.32 &  3 &  23.00 &  0.12 &  3 \\ 
L4j01    &  23.81 &  0.18 &  5 &  23.46 &  0.40 &  7 & \nodata & \nodata & .. \\ 
L4j02    &  23.35 &  0.09 &  5 &  22.59 &  0.13 &  7 &  22.37 &  0.12 &  3 \\ 
L4j03    &  23.87 &  0.13 &  5 &  23.06 &  0.18 &  7 &  22.91 &  0.15 &  3 \\ 
L4j05    &  23.55 &  0.10 &  6 &  22.66 &  0.12 &  6 &  22.46 &  0.14 &  3 \\ 
L4j06PD  &  22.13 &  0.04 &  4 &  21.70 &  0.03 &  4 &  21.74 &  0.04 &  2 \\ 
L4j07    &  22.96 &  0.19 &  5 &  22.10 &  0.09 &  3 &  21.91 &  0.04 &  2 \\ 
L4j08    &  23.51 &  0.16 &  5 &  22.79 &  0.20 &  7 &  22.43 &  0.32 &  3 \\ 
L4j10    &  23.75 &  0.21 &  3 &  23.06 &  0.12 &  2 &  22.82 &  0.06 &  3 \\ 
L4j11    &  23.55 &  0.32 &  8 &  23.04 &  0.30 &  5 &  22.77 &  0.26 &  3 \\ 
L4j12    &  23.61 &  0.08 &  4 &  23.14 &  0.11 &  3 & \nodata & \nodata & .. \\ 
L4k01    &  24.01 &  0.12 &  3 &  23.11 &  0.17 &  2 &  23.03 &  0.11 &  3 \\ 
L4k02    &  23.14 &  0.20 &  4 &  22.59 &  0.04 &  2 &  22.18 &  0.15 &  2 \\ 
L4k03    &  23.54 &  0.28 &  4 &  22.97 &  0.34 &  9 &  22.25 &  0.05 &  3 \\ 
L4k04    &  24.16 &  0.15 &  4 &  23.22 &  0.03 &  3 &  22.96 &  0.12 &  3 \\ 
L4k09    &  23.69 &  0.22 &  4 &  22.53 &  0.29 &  3 &  22.34 &  0.19 &  4 \\ 
L4k10    &  24.43 &  0.23 &  4 &  23.44 &  0.28 &  4 &  23.19 &  0.15 &  3 \\ 
L4k11    &  23.32 &  0.15 &  4 &  22.99 &  0.10 &  3 &  22.58 &  0.07 &  2 \\ 
L4k12    &  23.20 &  0.20 &  4 &  22.72 &  0.11 &  4 & \nodata & \nodata & .. \\ 
L4k13    &  23.99 &  0.16 &  3 &  23.15 &  0.12 &  3 &  23.03 &  0.17 &  3 \\ 
L4k14    &  24.08 &  0.14 &  4 &  23.32 &  0.19 &  4 &  22.93 &  0.95 &  3 \\ 
L4k15PD  &  23.22 &  0.07 &  4 &  22.48 &  0.17 &  3 &  22.12 &  0.02 &  2 \\ 
L4k16    &  23.99 &  0.15 &  4 &  22.93 &  1.27 &  5 &  23.32 &  0.12 &  2 \\ 
L4k17    &  23.07 &  0.17 &  4 &  22.55 &  0.12 &  5 &  22.33 &  0.05 &  3 \\ 
L4k18    &  23.61 &  0.09 &  4 &  23.08 &  0.34 &  3 & \nodata & \nodata & .. \\ 
L4k19    &  23.66 &  0.29 &  4 &  23.40 &  0.17 &  6 & \nodata & \nodata & .. \\ 
L4k20    &  23.80 &  0.27 &  4 &  23.19 &  0.29 &  3 &  22.47 &  0.46 &  2 \\ 
L4m01    &  23.83 &  0.14 &  5 & \nodata & \nodata & .. & \nodata & \nodata & .. \\ 
L4m02    &  23.43 &  0.25 &  8 &  22.72 &  0.10 &  3 & \nodata & \nodata & .. \\ 
L4m03    &  23.57 &  0.39 &  5 &  23.33 &  0.21 &  3 & \nodata & \nodata & .. \\ 
L4m04    &  23.58 &  0.53 &  5 & \nodata & \nodata & .. &  22.14 &  0.00 &  1 \\ 
L4n03    &  23.72 &  0.11 &  4 & \nodata & \nodata & .. &  22.72 &  0.49 &  3 \\ 
L4n04    &  23.60 &  0.17 &  4 &  22.45 &  0.11 &  4 &  22.50 &  0.00 &  1 \\ 
L4n05    &  23.69 &  0.24 &  4 & \nodata & \nodata & .. &  23.05 &  0.15 &  2 \\ 
L4n06    &  23.65 &  0.09 &  4 & \nodata & \nodata & .. &  23.49 &  0.79 &  3 \\ 
L4o01    &  22.98 &  0.09 &  4 &  22.21 &  0.39 &  3 &  21.95 &  0.76 &  3 \\ 
L4p01    &  23.90 &  0.13 &  3 &  23.48 &  0.41 &  3 & \nodata & \nodata & .. \\ 
L4p02    &  23.83 &  0.20 &  4 &  23.07 &  0.14 &  2 & \nodata & \nodata & .. \\ 
L4p03    &  23.17 &  0.11 &  4 &  22.45 &  0.14 &  4 &  22.09 &  0.11 &  2 \\ 
L4p04PD  &  21.96 &  0.14 &  4 &  22.33 &  1.43 &  4 &  21.58 &  0.09 &  3 \\ 
L4p05    &  23.57 &  0.08 &  4 &  22.74 &  0.08 &  6 &  22.38 &  0.16 &  3 \\ 
L4p06PD  &  22.34 &  0.10 &  4 &  21.81 &  0.23 &  4 &  21.56 &  0.30 &  3 \\ 
L4p07    &  22.32 &  0.34 &  4 & \nodata & \nodata & .. &  24.08 &  0.37 &  3 \\ 
L4p08PD  &  23.81 &  0.21 &  4 & \nodata & \nodata & .. &  22.87 &  0.14 &  3 \\ 
L4p09    &  23.61 &  0.21 &  4 &  22.88 &  0.17 &  7 &  22.62 &  0.14 &  2 \\ 
L4q03    &  23.57 &  0.07 &  3 &  23.01 &  0.09 &  7 & \nodata & \nodata & .. \\ 
L4q05    &  23.56 &  0.11 &  4 &  23.04 &  0.10 &  3 &  22.84 &  0.41 &  3 \\ 
L4q06    &  23.97 &  0.24 &  4 &  23.54 &  0.19 &  3 &  23.83 &  0.43 &  3 \\ 
L4q09    &  24.11 &  0.12 &  4 &  23.11 &  0.01 &  3 &  22.72 &  0.23 &  3 \\ 
L4q10    &  23.54 &  0.27 &  4 & \nodata & \nodata & .. &  22.44 &  0.03 &  3 \\ 
L4q11    &  23.92 &  0.14 &  4 & \nodata & \nodata & .. &  22.81 &  0.16 &  3 \\ 
L4q12PD  &  24.11 &  0.09 &  3 & \nodata & \nodata & .. & \nodata & \nodata & .. \\ 
L4q14    &  23.59 &  0.15 &  7 & \nodata & \nodata & .. &  22.89 &  0.34 &  3 \\ 
L4q15    &  24.04 &  0.24 &  4 & \nodata & \nodata & .. & \nodata & \nodata & .. \\ 
L4q16    &  23.71 &  0.17 &  4 & \nodata & \nodata & .. &  20.17 &  0.00 &  1 \\ 
L4v01    &  24.18 &  0.19 &  4 &  24.00 &  1.07 &  2 & \nodata & \nodata & .. \\ 
L4v02    &  23.80 &  0.02 &  4 &  23.01 &  0.27 &  3 & \nodata & \nodata & .. \\ 
L4v03    &  22.83 &  0.12 &  4 &  21.95 &  0.09 &  3 & \nodata & \nodata & .. \\ 
L4v04    &  24.13 &  0.09 &  4 &  23.32 &  0.22 &  3 &  23.45 &  0.23 &  2 \\ 
L4v05    &  24.11 &  0.17 &  4 &  23.32 &  0.12 &  2 &  23.04 &  0.07 &  2 \\ 
L4v06    &  23.66 &  0.15 &  4 &  23.03 &  0.11 &  3 &  22.51 &  0.12 &  2 \\ 
L4v08    &  23.95 &  0.16 &  4 &  23.17 &  0.27 &  2 &  23.16 &  0.02 &  2 \\ 
L4v09    &  23.52 &  0.08 &  5 &  23.03 &  0.08 &  4 &  22.83 &  0.09 &  3 \\ 
L4v10    &  23.87 &  0.21 &  4 &  22.93 &  0.20 &  3 &  22.85 &  0.05 &  3 \\ 
L4v11    &  24.16 &  0.20 &  4 &  23.64 &  0.77 &  4 &  23.58 &  0.17 &  3 \\ 
L4v12    &  24.00 &  0.08 &  4 &  23.44 &  0.20 &  3 &  23.60 &  0.13 &  4 \\ 
L4v13    &  22.72 &  0.09 &  4 &  22.68 &  0.70 &  5 &  22.32 &  0.02 &  3 \\ 
L4v14    &  23.25 &  0.16 &  4 &  22.57 &  0.17 &  5 & \nodata & \nodata & .. \\ 
L4v18    &  22.95 &  0.22 &  3 &  22.66 &  0.15 &  4 &  22.57 &  0.27 &  2 \\ 
l4h03    &  23.46 &  0.17 &  4 & \nodata & \nodata & .. & \nodata & \nodata & .. \\ 
l4h04    &  23.89 &  0.17 &  3 & \nodata & \nodata & .. &  23.48 &  0.21 &  3 \\ 
l4h17    &  23.59 &  0.09 &  4 & \nodata & \nodata & .. & \nodata & \nodata & .. \\ 
l4h19    &  23.91 &  0.39 &  4 & \nodata & \nodata & .. & \nodata & \nodata & .. \\ 
l4k05    &  23.74 &  0.23 &  4 & \nodata & \nodata & .. &  23.20 &  0.14 &  4 \\ 
l4k06    &  24.16 &  0.23 &  4 & \nodata & \nodata & .. &  23.96 &  0.32 &  2 \\ 
l4k08    &  23.94 &  0.70 &  4 & \nodata & \nodata & .. & \nodata & \nodata & .. \\ 
l4o02    &  23.79 &  0.37 &  5 &  23.53 &  0.38 &  3 & \nodata & \nodata & .. \\ 
l4q01    &  23.81 &  0.06 &  3 & \nodata & \nodata & .. & \nodata & \nodata & .. \\ 
l4q02    &  23.77 &  0.20 &  4 & \nodata & \nodata & .. & \nodata & \nodata & .. \\ 
l4q04    &  23.72 &  0.08 &  5 & \nodata & \nodata & .. & \nodata & \nodata & .. \\ 
l4q08    &  23.30 &  0.11 &  3 & \nodata & \nodata & .. & \nodata & \nodata & .. \\ 
l4v07    &  23.39 &  0.47 &  4 & \nodata & \nodata & .. & \nodata & \nodata & .. \\ 
l4v16    &  24.30 &  0.43 &  3 & \nodata & \nodata & .. & \nodata & \nodata & .. \\ 
l4v17    &  24.19 &  0.20 &  3 & \nodata & \nodata & .. & \nodata & \nodata & .. \\ 
l4v19    &  23.99 &  0.43 &  3 & \nodata & \nodata & .. & \nodata & \nodata & .. \\ 
U4j04PD  &  24.10 &  0.21 &  5 &  23.25 &  0.25 &  4 &  22.98 &  0.12 &  3 \\ 
U4j09    &  24.19 &  0.30 &  2 &  23.29 &  0.17 &  6 &  22.83 &  0.23 &  3 \\ 
U4n01    &  24.20 &  0.25 &  4 & \nodata & \nodata & .. &  22.80 &  0.00 &  1 \\ 
U4n02    &  24.13 &  0.32 &  4 & \nodata & \nodata & .. &  21.50 &  1.11 &  3 \\ 
u4h20    &  24.14 &  0.20 &  3 & \nodata & \nodata & .. &  23.13 &  0.13 &  3 \\ 
u4k07    &  24.61 &  0.25 &  3 & \nodata & \nodata & .. & \nodata & \nodata & .. \\ 
u4q13    &  24.31 &  0.11 &  3 & \nodata & \nodata & .. &  22.85 &  0.19 &  3 \\

%\loaddata{tables/L5PhotometryTable}
L5c02    &  23.59 &  0.13 &  2 &  22.65 &  0.07 &  4 &  22.42 &  0.06 &  3 \\ 
L5c03    &  23.96 &  0.27 &  3 &  23.27 &  0.05 &  3 &  23.01 &  0.15 &  3 \\ 
L5c06    &  24.27 &  0.22 &  5 & \nodata & \nodata & .. &  22.68 &  0.73 &  3 \\ 
L5c07PD  &  22.94 &  0.11 &  5 &  22.17 &  0.12 &  2 &  22.01 &  0.08 &  3 \\ 
L5c08    &  23.59 &  0.24 &  6 &  22.66 &  0.20 &  4 &  22.67 &  0.07 &  3 \\ 
L5c10PD  &  24.02 &  0.05 &  3 & \nodata & \nodata & .. &  23.10 &  0.11 &  3 \\ 
L5c11    &  23.66 &  0.47 &  4 &  23.16 &  0.14 &  3 &  23.10 &  0.18 &  3 \\ 
L5c12    &  22.28 &  0.04 &  4 & \nodata & \nodata & .. &  21.19 &  0.02 &  3 \\ 
L5c13PD  &  23.80 &  0.22 &  4 & \nodata & \nodata & .. &  23.16 &  0.12 &  2 \\ 
L5c14    &  23.41 &  0.09 &  3 & \nodata & \nodata & .. &  22.61 &  0.07 &  3 \\ 
L5c15    &  24.16 &  0.16 &  4 &  23.27 &  0.57 &  4 &  23.27 &  0.15 &  3 \\ 
L5c16    &  23.08 &  0.16 &  4 &  22.67 &  0.05 &  3 &  22.63 &  0.11 &  2 \\ 
L5c18    &  23.84 &  0.14 &  3 & \nodata & \nodata & .. & \nodata & \nodata & .. \\ 
L5c19PD  &  23.76 &  0.55 &  3 &  23.24 &  0.28 &  4 &  23.17 &  0.08 &  3 \\ 
L5c20PD  &  24.04 &  0.20 &  4 &  23.69 &  0.59 &  4 &  22.52 &  0.97 &  3 \\ 
L5c21PD  &  23.75 &  0.20 &  3 &  22.84 &  0.10 &  4 &  22.79 &  0.14 &  3 \\ 
L5c22    &  23.59 &  0.19 &  4 &  22.83 &  0.07 &  3 &  22.48 &  0.03 &  3 \\ 
L5c23    &  24.19 &  0.09 &  3 &  23.37 &  0.07 &  4 &  23.27 &  0.16 &  4 \\ 
L5c24PD  &  23.84 &  0.40 &  3 & \nodata & \nodata & .. &  22.86 &  0.90 &  4 \\ 
L5i01    &  23.66 &  0.27 &  4 & \nodata & \nodata & .. & \nodata & \nodata & .. \\ 
L5i02PD  &  23.85 &  0.16 &  4 & \nodata & \nodata & .. & \nodata & \nodata & .. \\ 
L5i03PD  &  23.77 &  0.21 &  4 &  22.97 &  0.58 &  3 & \nodata & \nodata & .. \\ 
L5i04    &  23.07 &  0.21 &  7 & \nodata & \nodata & .. & \nodata & \nodata & .. \\ 
L5i05    &  23.81 &  0.21 &  4 & \nodata & \nodata & .. & \nodata & \nodata & .. \\ 
L5i06PD  &  23.12 &  0.05 &  4 & \nodata & \nodata & .. &  22.41 &  0.20 &  3 \\ 
L5i08    &  23.20 &  0.61 &  4 & \nodata & \nodata & .. &  22.41 &  0.08 &  3 \\ 
L5j02    &  23.29 &  0.11 &  5 &  22.20 &  0.18 &  3 & \nodata & \nodata & .. \\ 
L5j03    &  23.16 &  0.07 &  3 &  22.50 &  0.10 &  3 &  22.23 &  0.05 &  4 \\ 
L5j04    &  22.51 &  0.11 &  5 &  21.64 &  0.18 &  3 &  21.52 &  0.15 &  2 \\ 
L5r01    &  23.65 &  0.22 &  4 & \nodata & \nodata & .. & \nodata & \nodata & .. \\ 
L5s01PD  &  20.84 &  0.02 &  3 & \nodata & \nodata & .. & \nodata & \nodata & .. \\ 
l5c01    &  24.05 &  0.55 &  5 & \nodata & \nodata & .. & \nodata & \nodata & .. \\ 
l5c04    &  24.21 &  0.33 &  6 & \nodata & \nodata & .. & \nodata & \nodata & .. \\ 
U5c17PD  &  24.18 &  0.30 &  3 & \nodata & \nodata & .. &  23.46 &  0.14 &  3 \\ 
U5j01PD  &  23.83 &  0.16 &  5 &  23.37 &  0.24 &  3 & \nodata & \nodata & .. \\ 
U5j06    &  23.67 &  0.23 &  5 &  23.11 &  0.25 &  4 &  23.06 &  0.16 &  3 \\ 
u5c09    &  24.35 &  0.21 &  2 & \nodata & \nodata & .. & \nodata & \nodata & .. \\ 
u5i07    &  24.29 &  0.41 &  4 & \nodata & \nodata & .. &  22.92 &  0.21 &  3 \\ 
u5j05    &  23.50 &  0.06 &  5 & \nodata & \nodata & .. &  22.83 &  0.09 &  2 \\

%\loaddata{tables/L7PhotometryTable}
L7a02    &  23.50 &  0.33 &  3 & \nodata & \nodata & .. &  22.74 &  0.13 &  4 \\ 
L7a03    &  23.80 &  0.07 &  4 &  23.26 &  0.12 &  4 &  23.27 &  0.15 &  3 \\ 
L7a04PD  &  23.29 &  0.11 &  4 & \nodata & \nodata & .. &  21.92 &  0.09 &  3 \\ 
L7a05    &  23.68 &  0.20 &  4 &  22.85 &  0.00 &  1 &  22.82 &  0.38 &  3 \\ 
L7a06    &  23.78 &  0.13 &  4 & \nodata & \nodata & .. &  22.92 &  0.12 &  4 \\ 
L7a07    &  23.43 &  0.20 &  4 &  22.75 &  0.11 &  3 &  22.55 &  0.11 &  4 \\ 
L7a10    &  23.63 &  0.04 &  4 &  23.35 &  0.22 &  4 &  23.26 &  0.11 &  3 \\ 
L7a11PD  &  23.34 &  0.16 &  5 &  22.64 &  0.10 &  4 &  22.67 &  0.10 &  4 \\ 
l7a12    &  23.86 &  0.13 &  6 & \nodata & \nodata & .. & \nodata & \nodata & .. \\ 
U7a01    &  24.13 &  0.16 &  4 &  23.29 &  0.26 &  4 &  22.94 &  0.27 &  2 \\ 
U7a08    &  24.00 &  0.18 &  4 &  23.62 &  0.14 &  4 &  23.25 &  0.32 &  3 \\ 
U7a09    &  24.11 &  0.21 &  3 &  23.45 &  0.16 &  4 &  23.41 &  0.28 &  3 \\

\enddata
\tablecomments
{'L' objects are the tracked, characterized (i.e. with flux above the 40\%
detection efficiency level) objects of CFEPS.
'l' objects are the non-tracked, characterized objects of CFEPS. There is
no ephemeris-based bias in those losses. Most of them were not recovered at
checkup (either too faint or sheared out of field coverage westward).
'U' objects are the tracked, non-characterized (i.e. with flux below 40\%
detection efficiency level) objects of CFEPS.
'u' objects are the non-tracked, non-characterized objects of CFEPS.
Magnitudes listed for photometric observations from CFHT.
Some numbers are missing because the corresponding object was not re-observed
in a particular filter from CFHT in photometric conditions. This is the case
for lost objects, some of the PD objects which we did not try to track, or
objects tracked solely at other facilities. g, r, i columns give the apparent
magnitude of the object in the correspondant filter. $\sigma_x$ is the
uncertainty on the magnitude in filter x. N$_x$ is the number of measurements
in filter x used to derive the apparent magnitude and its uncertainty.
}

\end{deluxetable}
}}

\section{Appendix C}
\label{sec:app_c}

Comparing previously-published population estimates of the main classical belt,
either as a whole or for the various components, with our present values must
ensure that the same limiting $H_g$ magnitude and the same region of the
phase space are adopted.
The main difference between P1 and the present L7 model is the higher 
$q$ cut-off that was applied to the P1 sample. 
Restricting our current sample to the same region of phase space as was 
modeled in P1 gives very similar population estimates for the main
belt. 
Secondly, P1's cold component was restricted in extend to $a \le 45$~AU and
hence had a smaller population than in our current model, and conversely, the
hot population was slightly overestimated compared to our current value, for
the region of phase space.
Lastly, because we use widely different $H$-magnitude slopes, the population
estimates should be compatible for the detected $H_g$=7--8 range,
but diverge for smaller TNOs (larger H). 
Scaling P1 to the $H_g$=8 limit, we find
$$
N_{\rm P1}({\rm H}_g \le 8.0) = 4400^{+1800}_{-1100},
$$
while restricting our currennt model to the same phase space gives
$$
N_{\rm All}({\rm H}_g \le 8.0) = 5800^{+1300}_{-1200},
$$
in reasonable agreement.

The latest independant population estimate of the main classical belt was done
by \citet{2001AJ....122..457T}, who estimate 38,000$^{+5400}_{-2700}$ objects 
bigger than $D = 100$~km, with uncertainties being $3\sigma$ confidence.
This number is more than 3 times smaller than our $H_g$=9.16 estimate. 
Can the two numbers be reconciled ?

First, one must match the size ranges of the population being estimated.
\citeauthor{2001AJ....122..457T} used a red albedo $p_R$=0.04 and a
solar red magnitude of -27.1. In this case a TNO of $D = 100$~km has an
absolute magnitude ${\rm H}_R = 8.8$. 
Assuming the same $\mpg$~-~R=0.8 color as we used in 
\citet{2009AJ....137.4917K},
this corresponds to ${\rm H}_g = 9.6$.
When looking at Fig.~9 of \citet{2001AJ....122..457T} one clearly sees that
either the assumption of an exponential luminosity function breaks at around
$m_R \sim 24$, or the debiasing is incorrect faintward of that value. In
particular, the lack of debiased objects fainter than 24 would push the
population estimate down. For the main belt model used by
\citeauthor{2001AJ....122..457T}, $m_R = 24$ corresponds to ${\rm H}_R \sim
7.7$ or ${\rm H}_g \sim 8.5$. 
Hence their population estimate is probably more applicable to
that limit but not to smaller sizes.
With $q = 4$, or $\alpha = 0.6$, the population estimates of
\citet{2001AJ....122..457T} is
$$
N_{\rm Trujillo}({\rm H}_g \le 8.5) = 8300^{+1200}_{-600}.  [3\sigma]
$$
Restricting our model to the same phase space and extrapolating our population
estimate out to H$_g \le 8.5$, we obtain
$$
N_{\rm All}({\rm H}_g \le 8.5) = 19,000^{+4100}_{-3700}, [95\% confidence]
$$
a factor of 2 larger than \citet{2001AJ....122..457T}
However, this last number is an extrapolation beyond the limit to which CFEPS
really measured the population. 
A more secure comparison from our point of view is done for H$_g \le
8.0$. The numbers become
$$
N_{\rm Trujillo}({\rm H}_g \le 8.0) = 4200^{+600}_{-300},
$$
and
$$
N_{\rm All}({\rm H}_g \le 8.0) = 5500^{+1300}_{-1100}. [95\% confidence]
$$
Hence we marginally agree with \citet{2001AJ....122..457T} at 
H$_g \le 8.0$.
At  H$_g > 8.0$, an extrapolation of our result (using our two
$H$ slopes determined at larger sizes) rapidly diverges from the
\citet{2001AJ....122..457T} estimate; if the slope does indeed
drop near $H\sim$8 (or $R\simeq$24) to a shallower slope
\citep{2008AJ....136...83F} then the two estimates are 
less discrepant.

\acknowledgements
\section*{Acknowledgements} This research was supported by funding from the
Natural Sciences and Engineering Research Council of Canada, the
Canadian Foundation for Innovation, the National Research Council of
Canada, and NASA Planetary Astronomy Program NNG04GI29G.  This project could
not have been a success without the dedicated staff of the Canada-France-Hawaii
telescope as well as the assistance of the skilled telescope operators at KPNO
and Mount Palomar.

We dedicate this paper to the memory of Brian G. Marsden, for his 
devotion to orbital determination and passionate encouragement to
observational planetary astronomers.

{\it Facilities:} In addition to \facility{CFHT (MegaPrime)} this work was made
possible thanks to access to facilities listed in Table~\ref{tab:followup}.

\eject
\newpage

\bibliographystyle{apj}
\bibliography{Petit}

\end{document}